\documentclass[useAMS,usenatbib,a4paper]{mn2e} 
 
\usepackage{graphics} 
\usepackage{graphicx} 
\usepackage{times}
\usepackage{amsmath} 
\usepackage{rotating}
 
\usepackage[usenames]{color} 

\usepackage{pdfpages}

\newcommand{\apj}{ApJ}
\newcommand{\apjl}{ApJ}

\newcommand{\aj}{AJ}
\newcommand{\pasp}{PASP}
\newcommand{\mnras}{MNRAS}
\newcommand{\araa}{ARA\&A}
\newcommand{\aap}{A\&A}
\newcommand{\nat}{Nature}
\newcommand{\aapr}{A\&A Rev.}
\newcommand{\aaps}{A\&AS}

\title[Fundamental properties of early-type stars]{Optical Interferometry of early-type stars  with PAVO@CHARA \\ \LARGE{I. Fundamental stellar properties}} 

\author[V. Maestro et al.]
{V. Maestro$^{1}$\thanks{E-mail: V.Maestro@physics.usyd.edu.au}, X. Che$^{2}$, D. Huber$^{1,3}$, M. J. Ireland$^{1,4,5}$, J. D. Monnier$^{2}$, T. R. White$^{1}$,
\newauthor   Y. Kok$^{1}$, J. G. Robertson$^{1}$, G. H. Schaefer$^{6}$, T. A. Ten Brummelaar$^{6}$, P. G. Tuthill$^{1}$\\
$^{1}$Sydney Institute for Astronomy, School of Physics, University of Sydney, NSW 2006, Australia \\
$^{2}$University of Michigan, Astronomy Department, 941 Denison Bldg, Ann Arbor, MI, USA \\
$^{3}$NASA Ames Research Center, Moffett Field, CA 94035, USA \\ 
$^{4}$Department of Physics and Astronomy, Macquarie University, NSW 2109, Australia \\
$^{5}$Australian Astronomical Observatory, PO Box 915, North Ryde, NSW 1670, Australia \\
$^{6}$Center for High Angular Resolution Astronomy, Georgia State University, PO Box 3969, Atlanta, GA 30302, USA}
\begin{document}
\maketitle
\date{\today}

\pagerange{\pageref{firstpage}--\pageref{lastpage}} \pubyear{2013}

\label{firstpage}

\begin{abstract}
We present interferometric observations of 7 main-sequence and 3 giant stars with spectral types from B2 to F6 using the PAVO beam combiner at the CHARA array.  We have directly determined the angular diameters for these objects with an average precision of 2.3\%. We have also computed bolometric fluxes using available photometry in the visible and infrared wavelengths, as well as space-based ultraviolet spectroscopy. Combined with precise \textit{Hipparcos} parallaxes, we have derived a set of fundamental stellar properties including linear radius, luminosity and effective temperature. Fitting the latter to computed isochrone models, we have inferred masses and ages of the stars. The effective temperatures obtained are in good agreement (at a 3\% level) with nearly-independent temperature estimations from spectroscopy. They validate recent sixth-order polynomial (B-V)-$T_\mathrm{eff}$ empirical relations \citep{Boyajian2012a}, but suggest that a more conservative third-order solution \citep{vanBelle2009} could adequately describe the (V-K)-$T_\mathrm{eff}$ relation for main-sequence stars of spectral type A0 and later. Finally, we have compared mass values obtained combining surface gravity with inferred stellar radius (\textit{gravity mass}) and as a result of the comparison of computed luminosity and temperature values with stellar evolutionary models (\textit{isochrone mass}). The strong discrepancy between isochrone and gravity mass obtained for one of the observed stars, $\gamma$\,Lyr, suggests that determination of the stellar atmosphere parameters  should be revised.

\end{abstract}

\begin{keywords}
Stars: early-type - Stars: fundamental parameters - Techniques: interferometric
\end{keywords}



\section{Introduction}
\label{sec:intro}  

Long-baseline optical interferometry (with baselines up to hundreds of meters in length) has enabled us to measure angular diameters of bright stars, with typical values of a few milliarcseconds. Combined with accurate parallax and photometry, these measurements allow direct determination of fundamental stellar properties, such as the linear radii of their photospheres or the effective surface temperature of the stars, and have become a very valuable tool to contrast observational results with stellar models of increasing complexity.

The pioneering work of \citet{Hanbury1974a}, using the Narrabri Stellar Intensity Interferometer (NSII), provided angular sizes of 32 O- to F- type stars. Subsequently, \citet{Code1976} established the empirical temperature scale for stars of spectral type F5 and earlier by means of combining \citet{Hanbury1974a} diameters with multiband spectra from which they inferred bolometric fluxes. The majority of the stars observed with the NSII belonged to luminosity classes I-III, and only about one third of them were main sequence or subgiant stars (luminosity classes IV or V), since the instrument favoured observations of stars with larger diameters, given the same surface brightness. For several decades, all the early type star (between O0 and A7) diameter measurements (a total of 16) came from the NSII observations \citep{Davis1997}. Even in more recent years, the papers referenced in the CHARM2 catalogue \citep{Richichi2005} contained just 24 entries corresponding to direct diameter measurements for main sequence or sub-giant stars. 
The advent of optical long baseline interferometry using hectometric baselines, particularly in the near infrared, has provided rapid progress in the number of main sequence and subgiant stars with direct diameter measurements. Most notably, \citet{vanBelle2009} reported interferometric diameter measurements for 44 G-type or later main sequence stars, deriving colour-temperature relations, and \citet{Boyajian2012a,Boyajian2012b} published results of a survey carried out on a combined sample of 77 dwarfs spanning from A to M spectral type, developing new empirical laws relating broad-band colours and effective temperature. 

As a result of resolution and sensitivity constraints particularly in the near-infrared, there is a clear sample bias, as only 5 out of the 121 stars studied in these papers belong to spectral class A or earlier. The major stumbling block has been that to access significant populations of hot stars, resolutions better than 1 milliarcsecond are required. The rise of beam combiners that can operate in the visible range of the spectrum with improved sensitivity, such as PAVO@CHARA, allows routine measurements of submilliarcsecond stellar diameters \citep{Huber2012b,White2013}, and offers the possibility to extend the spectral range to earlier type stars within similar sensitivity constraints.

Obtaining precise individual properties of B and A main sequence stars is of considerable importance in stellar astrophysics, since they represent the most massive and luminous objects that can be described by models containing simplifying assumptions such as LTE physics, hydrostatic equilibrium or purely radiative envelopes, in contrast to those corresponding to more massive or evolved objects, providing a useful benchmark for stellar atmosphere models. Furthermore, the intrinsic higher surface brightness of early type (mainly B and A) stars makes them suitable calibration stars \citep{Mozurkewich2003,Boden2003} for correction of instrumental and atmospheric effects in visible and near infrared interferometry. Therefore, direct measurement of submilliarcsecond angular stellar diameters has the potential to improve calibration of interferometric observations of objects with larger projected sizes.

The drawback is that stars earlier than F6 often rotate rapidly \citep{vanBelle2012}. As the rotational velocity approaches its critical value, the centrifugal force induces latitudinal temperature gradients \citep[an effect known as \textit{gravity darkening}, ][]{vonZeipel1924a,vonZeipel1924b,Maeder2000} and, depending on the inclination of the stellar rotation axis, the apparent stellar disk may look oblate. Projected rotational velocities can be used to model the effect of rotation in the interferometric observables \citep{Yoon2007}, with the finding that effects of low rotation rates can be  neglected for low spatial frequencies.

In this paper, we present results of our pilot study on a small sample of 10 nearby stars with submilliarcsecond angular size with spectral types in the B2-F6 range. Computed bolometric fluxes, together with precise \textit{Hipparcos} parallaxes have been used to derive the fundamental stellar parameters linear radius, luminosity and effective temperature , with results that are in good agreement with measurements obtained using independent methods. Finally, we have estimated stellar mass and ages by means of isochrone model fitting.


\section{Target sample and observations} 
\label{sec:observations}

\subsection{Target sample}\label{subsec:targetsample}

The stars observed in our study are extensively used as calibrators in near-infrared interferometry, mainly in observations using the MIRC (Michigan InfraRed Combiner) instrument \citep[see, e.g.][]{Monnier2007a,Zhao2009,Che2011}. In addition to their use in fundamental parameters of the observed stars (in combination with other measurements), angular diameter estimation constitutes a direct measurement of the interferometric response independent of diameter estimates based on indirect methods \citep[see][and references therein]{Cruzalebes2010} that rely on high fidelity SED templates or stellar atmosphere models. 

Table~\ref{tab:params} lists the objects observed, as well as the physical parameters describing their stellar atmospheres ($T_\mathrm{eff}$, $\log g$ and metallicity), determined from spectroscopic and photometric observations. It also includes the measured \textit{Hipparcos} parallaxes \citep{vanLeeuwen2007}.  These parallaxes correspond to distances smaller than $\sim$200\,pc, with uncertainties ranging from 1\% to 5\%  for most of the stars observed. The more distant stars $\sigma$\,Cyg and $o$\,And have larger 17\% and 11\% errors at a distance of 880 and 210 pc respectively.


\begin{table*}
\caption{ Physical parameters of the stars presented in this work. The objects are separated into main-sequence and sub giant stars (\textit{top}) and giants and supergiants (\textit{bottom}) in ascending temperature order.}
\label{tab:params}
\centering
\includegraphics[angle=90,height=190mm]{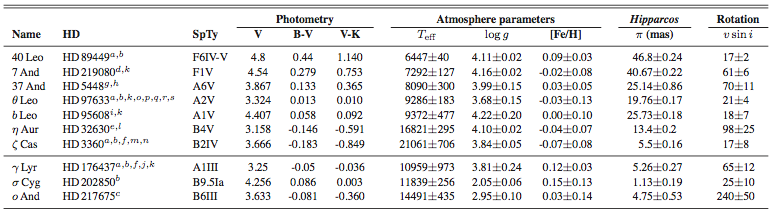}
\vfill
\textbf{Notes.} $V$ magnitude and $(B-V)$ colour taken from references in SIMBAD. $(V-K)$ colour computed using $K$-band photometry taken from the General Catalogue of Photometric Data \citep[GCPD][]{Mermilliod1997}. Stellar atmosphere parameters for every object have  been averaged over the values presented in$^a$\citet{Prugniel2011},$^b$\citet{Wu2011},$^c$\citet{Pier2003},$^d$\citet{Erspamer2003}, $^e$\citet{Fitzpatrick2005}, $^f$\citet{Koleva2012},$^g$\citet{Blackwell1998},$^h$\citet{Gardiner1999},$^i$\citet{AllendePrieto1999},$^j$\citet{Balachandran1986},$^k$\citet{Ammons2006},$^l$\citet{Adelman2002},$^m$\citet{Gies1992},$^n$\citet{Nieva2012},$^o$\citet{Adelman1986},$^p$\citet{Adelman1988},$^q$\citet{Cottrell1986},$^r$\citet{Hill1993},$^s$\citet{Smith1993}. Parallaxes from \citet{vanLeeuwen2007}. Projected rotational velocities ($v\sin i$) taken from \citet{Glebocki2005}.
\end{table*}


The stars in our sample span from B2 to F6 spectral types (effective temperatures in the 21000-6400\,K range). Most of them (seven) are main sequence or subgiants (luminosity classes V or IV). Two are giants ($\gamma$\,Lyr, $o$\,And), and one is a supergiant star ($\sigma$\,Cyg). Main-sequence stars of spectral types earlier than $\sim$F6 ($M>1.5M_{\odot}$), exhibiting radiative envelopes, are expected to rotate rapidly \citep[see][and references therein]{vanBelle2012}. As a consequence, they can show significant projected oblateness depending on the inclination of their polar axis. All the stars in our sample, with the sole exception of $o$\,And \citep[$v\sin i \sim 250$\,km/s, see][]{Balona1999,Glebocki2005}, have projected rotational velocities that are significantly less than 50\% of the critical rotational velocity \citep[see typical values for different spectral types in ][]{Tassoul2000} , and therefore no important deviations from projected circular shapes are expected \citep{Fremat2005}. For $o$\,And, \citet{Clark2003} reported that $o$\,And could be seen nearly equator-on, implying that the star rotates at $\sim$50\% of the critical velocity,  which would result in an equatorial radius less than 4\% larger than the polar radius \citep{Owocki1994}. Unfortunately, our observations are not sensitive to oblateness, given that we observed this object using only one baseline.

Among the stars studied, three objects ($\sigma$\,Cyg, $b$\,Leo and $o$\,And) present some hints for the existence of companions at less than 1 degree in separation according to the \citet{Eggleton2008} compilation. Nevertheless, only one object ($o$\,And) shows observational evidence of the existence of close companions according to the Washington Double Star catalogue \citep[WDS][]{Mason2001}, the  Multiple Star Catalogue \citep[MSC][]{Tokovinin1997} and the 9$^{th}$ Catalogue of Spectroscopic Binary Orbits \citep[SB9][]{Pourbaix2004}. $o$\,And is a complex object, consisting of two components A and B\citep{Olevic2006}, separated by 0.34\,arcseconds (m$_V(A)$=3.63; m$_V(B)$=6.03). Both components have been described as spectroscopic binaries. The Aa-Ab components of the main spectroscopic binary have an estimated separation of 0.05\,arcseconds. The brightest star of the pair is believed to $\gamma$\,Cas type variable that injects material in a circumstellar shell in rapid discrete ejections \citep{Clark2003}, switching between Be- and B-type spectra in a timescale of $\sim10^3$\,days. The short-term photometric variability of this object does not correlate with an enhancement in the shell emission and seems to be photosperic in origin. All these features make  $o$\,And an extraordinarily difficult object to study, with conditions that might be relatively far from the simplified assumptions (circular projected shape, isothermal surface) considered throughout this paper; its continued use as an interferometric calibrator is not advised. Nevertheless, the measured angular diameter and effective temperature of $o$\,And show that the influence of companion objects is smaller than the precision of  our observations.

\subsection{Interferometry}\label{subsec:observations}
Interferometric observations of our target sample were carried out using the PAVO (\textit{Precision Astronomical Visible Observations}) beam combiner \citep{Ireland2008}, located at the CHARA Array \citep{tenBrummelaar2005} on Mt Wilson Observatory (California, USA). The CHARA (\textit{Center for High Angular Resolution Astronomy}) Array is an optical interferometer, consisting of six 1-m telescopes arranged in a Y-shaped configuration. Operating in visible and infrared wavelengths, it provides a total of 15 different baselines at different orientations with lengths in the range 34-331\,m. With the longest operational baselines available in the world provided by the CHARA array, and the use of visible light (0.6-0.9\,$\mu$m) in PAVO, the instrumentation used in this study delivers the highest angular resolution yet achieved ($\sim$0.3\,mas in the visible).

The PAVO instrument  \citep[see detailed description in][]{Ireland2008} is a pupil-plane Fizeau beam combiner optimised for sensitivity and high angular resolution. We briefly summarise the basics of the instrument here. Visible and infrared light are separated by a dichroic with a cutoff at 1$\mu$m. The visible beams (up to three) enter PAVO, and are focused by a set of achromatic lenses in an image plane. The beams are passed through a 3-hole non-redundant mask that acts as a spatial filter. After going through the mask, the beams interfere and produce spatially modulated pupil-plane fringes. The fringes are formed on a lenslet array that divides the pupil in 16 independent segments, allowing an optimal usage of the multi-r$_0$ apertures of the CHARA array. Finally, a prism disperses the fringes and these are re-imaged and recorded on a low-noise readout EMCCD detector. Early PAVO@CHARA results have been presented by \citet{Bazot2011}, \citet{Derekas2011} and \citet{Huber2012b,Huber2012a}.

Observations of the objects listed in Table~\ref{tab:params} using PAVO@CHARA were carried out in July 2010 (2$^{\mathrm{nd}}$-3$^{\mathrm{rd}}$), May 2011 (12$^{\mathrm{th}}$-13$^{\mathrm{th}}$),  August 2011 (11$^{\mathrm{th}}$), October 2011 (2$^{\mathrm{nd}}$-3$^{\mathrm{rd}}$), August 2012 (5$^{\mathrm{th}}$-6$^{\mathrm{th}}$)  and September 2012 (7$^{\mathrm{th}}$). Most of the observations  were done using one baseline (two telescopes) at a time, with the only exception of $\sigma$\,Cyg, that has also been observed in three-telescope (using the S2E2W2 triangle) mode, allowing simultaneous data collection in three baselines. The baselines used are given in Table~\ref{tab:baselines}. Raw interferometric data ($V^2$) were obtained through the use of standard procedures for PAVO@CHARA data  \citep{Ireland2008,Maestro2012}. Some of the objects ($\zeta$\,Cas, 37\,And and $b$\,Leo) have been observed on only one night.


\begin{table}\caption{CHARA baselines used. Position angle (PA) is measured in degrees East of North.}
\label{tab:baselines}
\begin{center}
\begin{tabular}{lcr}
  \hline
  \hline
\textbf{Baseline} & \textbf{Length (m)} & \textbf{PA (deg)} \\ 
  \hline
 W2W1 & 107.93 & -80.9 \\ 
 W2E2 & 156.26 & +63.2 \\ 
 W2S2 & 177.45 & -20.9 \\ 
 W2S1 & 210.98 & -19.1 \\ 
 E2S2 & 248.13 & +17.9 \\ 
 W1S2 & 249.39 & -42.8 \\ 
 W1E2 & 251.34 & +77.7 \\ 
 E1S1 & 330.70 & +22.3 \\ 
   \hline
\end{tabular}
\end{center}
\end{table}

Gauging the point-source response of the interferometer is essential to obtain an accurate calibration of the observed sources. To this end, and according to standard practice in optical interferometry \citep{Boden2003,vanBelle2005}, we interleave observations of the science targets with others of calibration stars, in such manner that a bracket calibrator-target-calibrator is completed in 15-20 minutes. The calibration stars match, as closely as possible, the ideal point-like ($\theta< 0.25$\,mas) source located at the smallest angular distance in the sky from the object (at an average distance of $\sim$7.5 degrees between target and calibrator). Table~\ref{tab:calibrators} shows all the calibrators used in our study. Expected diameters have been computed using $V-K$ colours \citep{Kervella2004}, dereddened according to the interstellar extinction maps presented by \citet{Drimmel2003}. We have checked each calibrator in the literature for possible multiplicity or variability prior to observations. Analysis of the data obtained for HD\,216523 revealed that the object is in fact a binary star, and therefore it was excluded from the list of calibrators. PAVO@CHARA sensitivity limits impose a selection bias on the calibration sources, favouring the use of distant late B to early A-type stars as calibrators, which are prone to show fast rotation and therefore non circular projected shapes \citep[][]{DomicianoDeSouza2002,vanBelle2012} . We circumvent this issue by choosing calibrators with low projected rotational velocities, $v\sin i$, or accounting for an increased uncertainty in the predicted diameter that reflects deviations from the spherical shape (see Section~\ref{sec:results}).  To account for some small correlated instrument systematics that persist after calibration using unresolved sources when the star has been observed only one night, we have included and additional 5\% uncertainty to the visibility-squared ($V^2$) errors measured, based on repeated observations of the same  object during  several nights.



\begin{table}\caption{List of calibration stars used, including the spectral type, $V-K$ colour and extinction $E(B-V)$, the predicted diameter (in milliarcseconds) from $V-K$ using prescriptions of \citet{Kervella2004}, as well as projected rotational velocities ($v\sin i$, in km/s) from \citet{Glebocki2005}. Photometry has been taken from references in SIMBAD database and the General Catalogue of Photometric Data \citep[GCPD][]{Mermilliod1997}.}
\label{tab:calibrators}
\begin{center}
\begin{tabular}{llrccr@{$\pm$} lr}
\hline
\hline
\textbf{HD} & \textbf{SpTy} & \textbf{$V-K$} & \textbf{$E(B-V)$} & \textbf{$\theta_{V-K}$} & \multicolumn{2}{c}{\textbf{$v\sin i$}} & \textbf{ID} \\
\hline
HD\,1279 & B7III & -0.157 & 0.037 & 0.202 & 25 & 9 & $^{bcg}$ \\
HD\,1606 & B7V & -0.358 & 0.037 & 0.161 & 120 & 22 & $^{c}$ \\
HD\,4142 & B5V & -0.334 & 0.034 & 0.193 & 160 & 24 & $^{cg}$ \\
HD\,10390 & B9V & -0.147 & 0.007 & 0.180 & 58 & 15 & $^{c}$ \\
HD\,29526 & A0V & -0.001 & 0.017 & 0.243 & 80 & 13 & $^f$ \\
HD\,29721 & B9III & 0.198 & 0.037 & 0.242 & 232 & 29 & $^f$ \\
HD\,88737 & F9V & 1.334 & 0.006 & 0.459 & 10 & 3 & $^{a}$ \\
HD\,89363 & A0 & 0.117 & 0.015 & 0.150 &  \multicolumn{2}{c}{-} & $^{ae}$ \\
HD\,92825 & A3V & 0.155 & 0.008 & 0.352 & 188 & 18 & $^{de}$ \\
HD\,93702 & A2V & 0.239 & 0.014 & 0.326 & 208 & 15 & $^{de}$ \\
HD\,171301 & B8IV & -0.180 & 0.025 & 0.184 & 53 & 15 & $^{h}$ \\
HD\,174262 & A1V & 0.074 & 0.025 & 0.226 & 111 & 21 & $^h$ \\
HD\,174567 & A0V & 0.073 & 0.044 & 0.160 & 18 & 7 & $^h$ \\
HD\,176871 & B5V & -0.162 & 0.038 & 0.210 & 268 & 34 & $^h$ \\
HD\,179527 & B8III & -0.074 & 0.046 & 0.208 & 30 & 12 & $^h$ \\
HD\,197392 & B8III & -0.21 & 0.212 & 0.213 & 33 & 9 & $^i$ \\
HD\,204403 & A5V & -0.489 & 0.027 & 0.227 & 117 & 18 & $^i$ \\
HD\,207516 & B8V & -0.157 & 0.016 & 0.179 & 103 & 18 & $^i$ \\
HD\,211211 & A2V & 0.061 & 0.018 & 0.243 & 238 & 27 & $^{ij}$ \\
HD\,219290 & A0V & -0.011 & 0.023 & 0.178 & 50 & 9 & $^{bj}$ \\
HD\,222304 & B9V & -0.052 & 0.024 & 0.269 & 165 & 25 & $^b$ \\
\hline
\end{tabular}
\end{center}
\textbf{Notes.} Calibrator not used: HD\,216523 (binary star). The last column refers to the ID of the target star for which the calibrator has been used (see first column of Table~\ref{tab:obsdiams}).
\end{table}


\section{Fundamental stellar parameters} 
\label{sec:results}

\subsection{Angular diameters}\label{ssec:angulardiameters}

\begin{figure*}
  \centering  
  \includegraphics[width=0.99\linewidth]{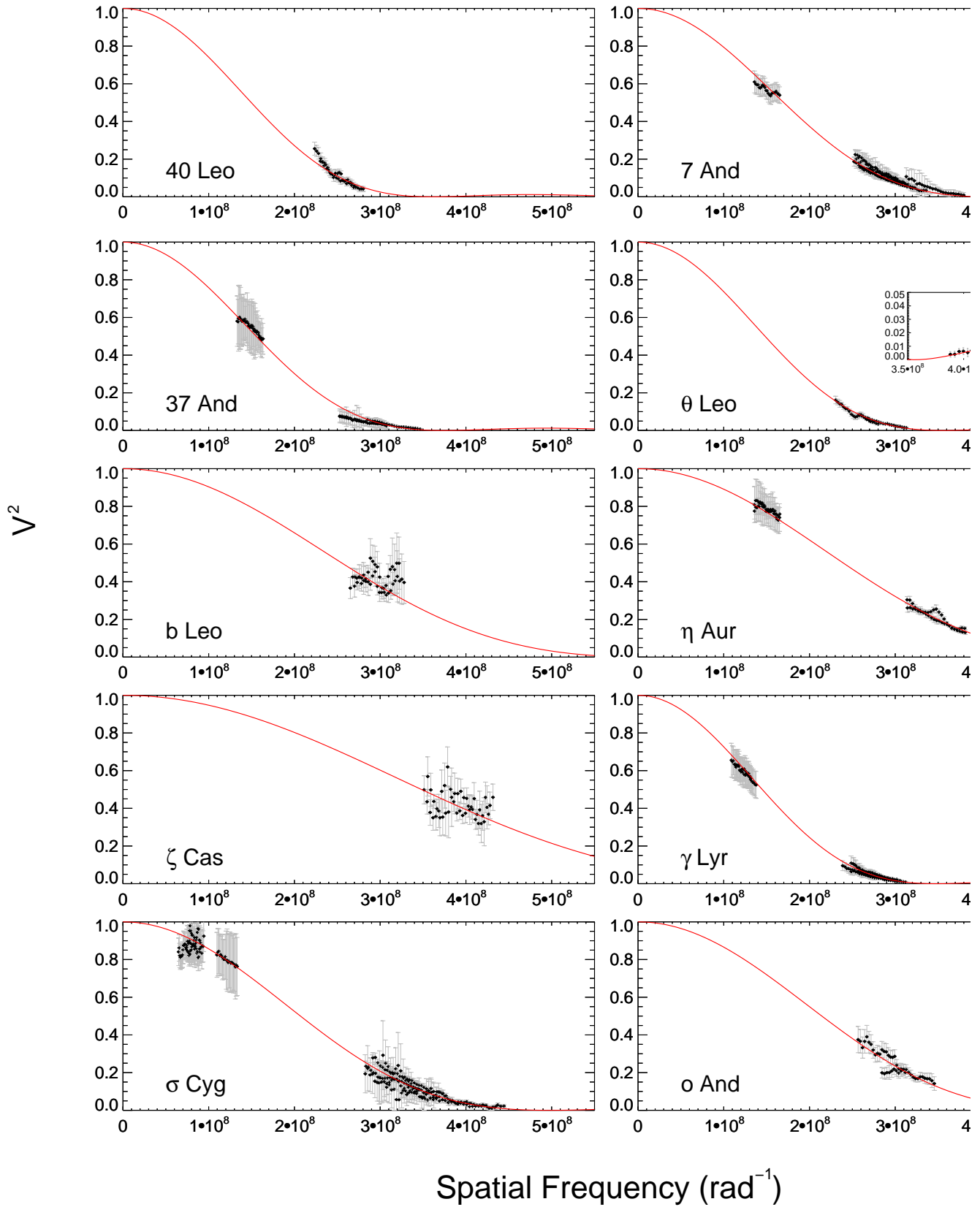}
  \caption{Squared visibility vs. spatial frequency (defined as projected baseline divided by wavelength) for all stars in our sample. Red solid lines show the fitted limb-darkened disk model. Error bars for each star have been scaled so that the reduced-$\chi^2$ equals unity.
  }
  \label{fig:v2plots}
\end{figure*}

The squared-visibility ($V^2$) measurements obtained for each of the stars presented in this work were fitted to the single star limb-darkened disk model \citep[given by ][]{Hanbury1974b}

\begin{align}\label{eq:lddiam}
	V=\left(\frac{1-\mu_\lambda}{2}+\frac{\mu_\lambda}{3}\right)^{-1} \times \notag \mspace{150mu}\\
	\left[(1-\mu_\lambda)\frac{J_1(x)}{x} + \mu_\lambda\sqrt{\frac{\pi}{2}}\frac{J_{3/2}(x)}{x^{3/2}}\right]
\end{align}
\\
with

\begin{equation}\label{eq:arglddiam}
	x=\frac{\pi B \theta_{\mathrm{LD}}}{\lambda}
\end{equation}

\noindent where $V$ is the visibility, $\mu_\lambda$ is the linear limb-darkening coefficient, $J_n(x)$ is the $n$th-order Bessel function, $B$ is the projected baseline,  $\theta_{\mathrm{LD}}$ is the angular diameter after limb-darkening correction, and $\lambda$ is the wavelength at which the stars are observed. We use R-band linear limb-darkening coefficients interpolating within the model grid of \citet{Claret2011}, using the atmosphere parameters given in Table~\ref{tab:params}, and assuming a microturbulent velocity of 2 km/s. The value assumed for $\mu_\lambda$ and its error are the result of taking the median and the  0.158 and 0.842 quantiles of the interpolated values corresponding to 10$^3$ realisations of the normally distributed values of the stellar atmosphere parameters $T_\mathrm{eff}$, $\log g$ and [Fe/H]. The uniform disc diameters were obtained by simply assuming $\mu_\lambda=0$ in Equation~\ref{eq:lddiam}. Figure~\ref{fig:v2plots} displays the limb-darkening disc model fit to the calibrated $V^2$. The observations made, as well as the uniform disc (computed assuming $\mu_R$=0 in equation~\ref{eq:lddiam}) and limb-darkened disc diameters estimated, are summarised in Table~\ref{tab:obsdiams}. 


\begin{table*}\caption{Summary of observations included in this work and measured angular diameters.}
\label{tab:obsdiams}
\begin{tabular}{rrlrp{1cm}r@{$\pm$}l r@{$\pm$}l r@{$\pm$}l r@{$\pm$}l }
\hline
\hline
\textbf{ID} & \textbf{Name} & \textbf{HD} & \textbf{N$_{V^2}$} & \textbf{Baselines} & \multicolumn{2}{c}{\textbf{$\mu_R$}} & \multicolumn{2}{c}{\textbf{$\theta_{UD}$ (mas)}} & \multicolumn{2}{c}{\textbf{$\theta_{LD}$ (mas)}} & \multicolumn{2}{c}{\textbf{$\theta_{V-K}$ (mas)}} \\
\hline
$^a$      &  40\,Leo                   &  HD\,89449      & 46 &  E2W1                                                & 0.48 & 0.04 & 0.706 & 0.026 & 0.731 & 0.030 & 0.747 & 0.012 \\
$^b$      & 7\,And  		        &  HD\,219080  & 207 &  W1W2, E2W1, S1W2, S2W1   & 0.44 & 0.04 & 0.622 & 0.006 & 0.648 & 0.008 & 0.649 & 0.013 \\
$^c$      &  37 And                      &  HD\,5448        & 115 &  W1W2, E2W1, S2W1           & 0.43 & 0.03 & 0.678 & 0.012 & 0.708 & 0.013 & 0.692 & 0.013 \\
$^d$      &  $\theta$\,Leo      &  HD\,97633      & 69 &  E2W1, S1E1                                   & 0.39 & 0.03 & 0.710 & 0.023 & 0.740 & 0.024 & 0.721 & 0.017 \\
$^e$      &  b\,Leo                       &  HD\,95608      & 46 &  E2W1                                                 & 0.38 & 0.03 & 0.416 & 0.016 & 0.430 & 0.017 & 0.456 & 0.009 \\
$^f$      &  $\eta$\,Aur            &  HD\,32630     & 115 &  W1W2, S2W1                         & 0.26 & 0.03 & 0.444 & 0.011 & 0.453 & 0.012 & 0.539 & 0.010 \\
$^g$      &  $\zeta$\,Cas           &  HD\,3360        & 46 &  S1E1                                               & 0.26 & 0.03 & 0.305 & 0.010 & 0.311 & 0.010 & 0.376 & 0.009 \\
\hline
$^h$      &  $\gamma$\,Lyr   &  HD\,176437  & 161 &  W1W2, W1E2, S1W2         & 0.34 & 0.03 & 0.729 & 0.008 & 0.753 & 0.009 & 0.738 & 0.016 \\
$^i$      &  $\sigma$\,Cyg     &  HD\,202850  & 368 &  W1W2, S2W1, S2E2W2    & 0.35 & 0.04 & 0.511 & 0.014 & 0.527 & 0.016 & 0.588 & 0.013 \\
$^j$      &  $o$\,And                 &  HD\,217675  & 46 &  E2W1                                                & 0.31 & 0.03 & 0.494 & 0.012 & 0.508 & 0.015 & 0.526 & 0.011 \\
\hline
\end{tabular}
\end{table*}


Errors in diameter have been estimated through model fitting of Equation~\ref{eq:lddiam} using synthetic datasets \citep{Derekas2011,Huber2012b}. These datasets are generated considering the uncertainties in: \textit{(1)} the measured $V^2$ values for target and calibrators, \textit{(2)} the adopted PAVO wavelength scale ($\pm$4.5\,nm), \textit{(3)} the calibrator angular sizes ($\pm$5\%, except those cases where the calibrator is expected to show larger projected oblateness), and \textit{(4)} linear limb-darkening coefficient (see Table~\ref{tab:obsdiams}). All quantities are assumed to have values that are normally distributed, using 2$\cdot$10$^4$ simulated datasets for each diameter estimation. Possible correlation between adjacent wavelength channels is also taken into account. The median and the width (derived from the 0.158 and 0.842 quantiles) of the distribution of the fitted diameters give the adopted values for the measured diameter and its uncertainty. The results have been adjusted to assume a reduced-$\chi^2=1$, compensating for underestimation of the squared-visibility error estimates \citep{Berger2006}.

\begin{figure}
  \centering  
  \includegraphics[width=1\linewidth,angle=0]{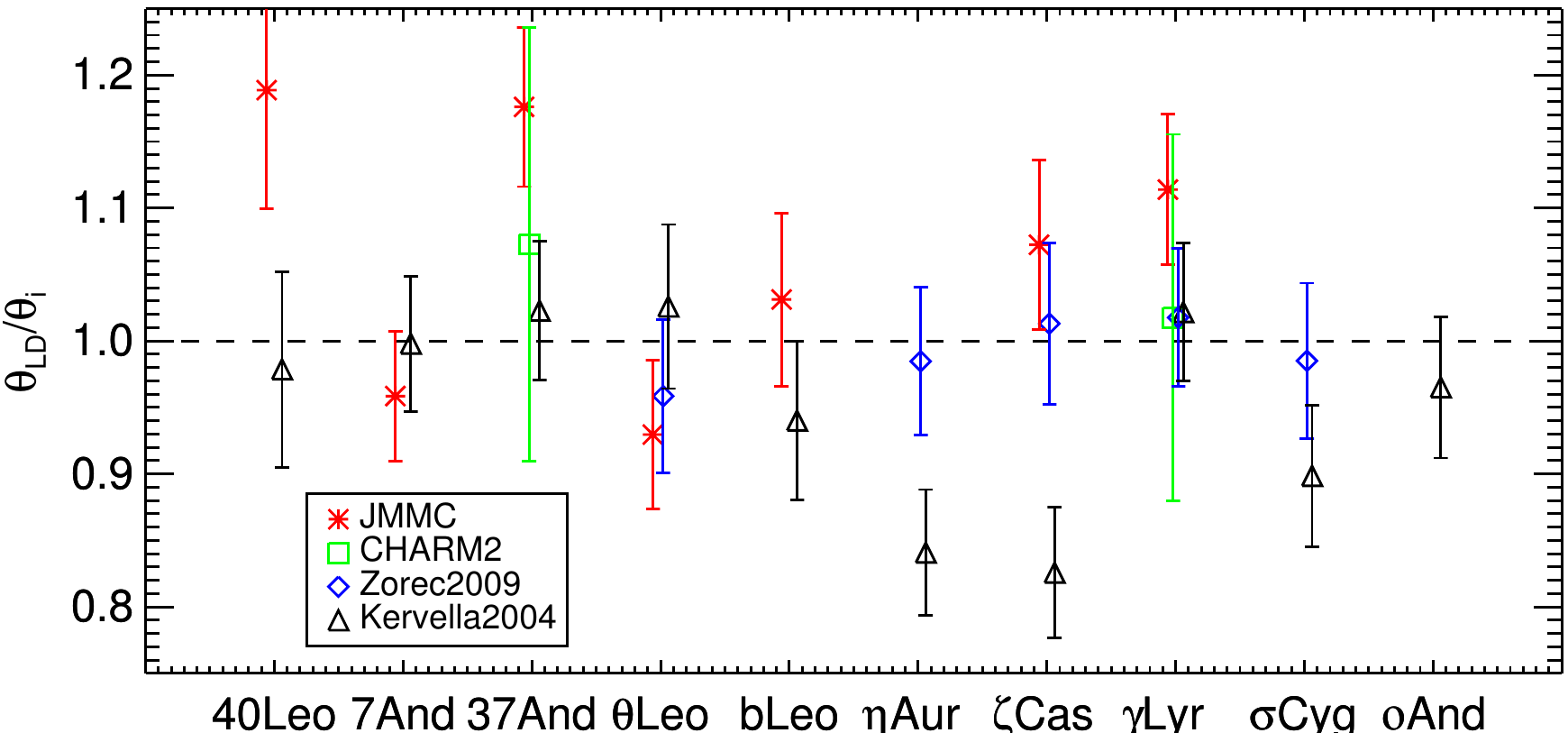}
  \caption{Fractional differences $\theta_{LD}/\theta_{i}$ between angular diameters measured with PAVO and diameters determined using color-surface brightness relation \citep[JMMC catalogue and Kervella relation][]{Lafrasse2010, Kervella2004}, SED fits \citep{Zorec2009} and direct interferometric measurements collected in the CHARM2 catalogue \citep{Richichi2005}.}
  \label{fig:compdiam}
\end{figure}

Derived angular diameters show an average precision of 2.3\%. Figure~\ref{fig:compdiam} shows the comparison between the measured limb-darkened diameters and different estimations using long baseline interferometry  \citep[CHARM2 meta-catalogue][]{Harmanec1996,Lane2001,Vakili1997}, or indirect methods: surface brightness methods and calibrations using color indices\footnote{\tt{http://cdsarc.u-strasbg.fr/viz-bin/Cat?II/300}} \citep[JMMC catalogue][]{Lafrasse2010,Kervella2004}, and using detailed spectrophotometry in the visible and ultraviolet \citep{Zorec2009}. The overall agreement with other results is good (the average value of  $\theta_{\mathrm{LD}}/\theta_{i}$ is 0.99 with a scatter of $\pm$0.06). $V^2$ data for $o$\,And does not show evidence of departure from circular shape, although this could be due to the similar orientation, projected on sky, of the baselines used. Stars of earlier spectral types , and particularly $\sigma$\,Cyg, $\eta$\,Aur and $\zeta$\,Cas, depart significantly from the range of spectral types of the stars used to infer Kervella's $(V-K)$-diameter relation (A- to K-type), and therefore show the larger deviations. It is worth noting the good concordance between observations and predictions made by \citet{Zorec2009}.

Second lobe $V^2$ measurements of $\theta$\,Leo make possible simultaneous estimation of both diameter and linear limb-darkening coefficient $\mu$ for this object. As a result, we obtain $\theta(\theta\,\textrm{Leo})$= 0.747$\pm$0.024 and $\mu(\theta\,\textrm{Leo})$=0.47$\pm$0.03. This constitutes a modest 1\% increment in the estimated diameter, but a remarkable increase in the limb-darkening effect with respect to the fixed $\mu$ estimation ($\theta(\theta\,\textrm{Leo})$= 0.740$\pm$0.024; $\mu(\theta\,\textrm{Leo})$=0.39$\pm$0.03). Given the spectral type  (A2V), and the low projected rotational velocity \citep[$v\sin i \simeq 25$\,km/s][]{Royer2007}, $\theta$\,Leo is likely to be a fast rotating star viewed nearly pole on. This results in a higher than expected drop in intensity near the border of the apparent stellar disc, due to the alignment of limb-darkening and gravity darkening, similar to that observed in other objects \citep[for example, the widely studied case of the A0V fast rotator Vega][]{Peterson2006,Aufdenberg2006,Monnier2012}.

\subsection{Bolometric fluxes}\label{ssec:bolflux}

We have computed the bolometric flux, $F_{bol}$, for all the targets in our sample by fitting the observed absolute spectral energy distribution (SED) to grids of ATLAS9 model atmospheres of solar metallicity computed by \citet{Castelli2003}. The fit requires collection of available photometry in the Hipparcos ($H_p$ bandpass), Tycho ($B_tV_t$), Johnson ($UBVRIJHK$), Geneva ($UBB1B2V$), WBVR (WBVR) and Stromgren ($ubvy$) phootometric systems. For the bright objects in our sample, \textit{2MASS} $JHK_s$ photometric measurements are generally saturated, and therefore are not suitable for our purpose, with the sole exception of $b$\,Leo, for which no other near-infrared colours were found. Photometry longward of  near-infrared passbands was not included, as they are frequently affected by infrared excess with non-stellar origin. 

Half the stars in the sample have spectral types earlier than A0, implying surface temperatures in excess of 10000\,K. Therefore these objects radiate predominantly in the ultraviolet. In order to improve the fit of the emergent flux in the ultraviolet region, we have included co-added low-resolution flux-calibrated ultraviolet spectra (in the 1150-1980\,{\AA} and 1850-3350\,{\AA} ranges) retrieved from the \textit{International Ultraviolet Explorer (IUE)} archive. To avoid occasional flux miscalibration close to the edges of the spectral range covered, we have considered only the 1200-1900\,{\AA} and 1900-3300\,{\AA} regions.

All photometric data have been corrected from interstellar reddening using the maps of \citet{Drimmel2003} and the extinction description presented in \citet{Fitzpatrick1999}. In the particular case of the highly reddened $\sigma$\,Cyg, we have used the individual interstellar extinction curved presented in \citet{Wegner2002}. Photometry was calibrated in flux \citep[using filter responses and zero points from][]{Kornilov1996,Bessell1998,Gray1998,Cohen2003,Bessell2012} and subsequently fitted to the grid of theoretical line-blanketed ATLAS9 spectra, interpolating in both $T^{model}_\textrm{eff}$ and $\log\,g$ model parameters. The code can also estimate the reddening correction, yielding to $E(B-V)$ values similar to those mentioned above. Table~\ref{tab:bolflux} shows the estimated bolometric fluxes, $F_{bol}$, computed as the numerical integral of the theoretical spectrum corresponding to the best model fit parameters $T_\textrm{eff}^{model}$ and $\log\,g$ (also listed in Table~\ref{tab:bolflux}). Uncertainties in the estimated parameters are computed using a synthetic population of 10$^3$ datasets accounting for errors (correlated and uncorrelated) in the photometry used and in its calibration in absolute flux.

\begin{figure*}
  \centering  
  \includegraphics[width=0.99\linewidth]{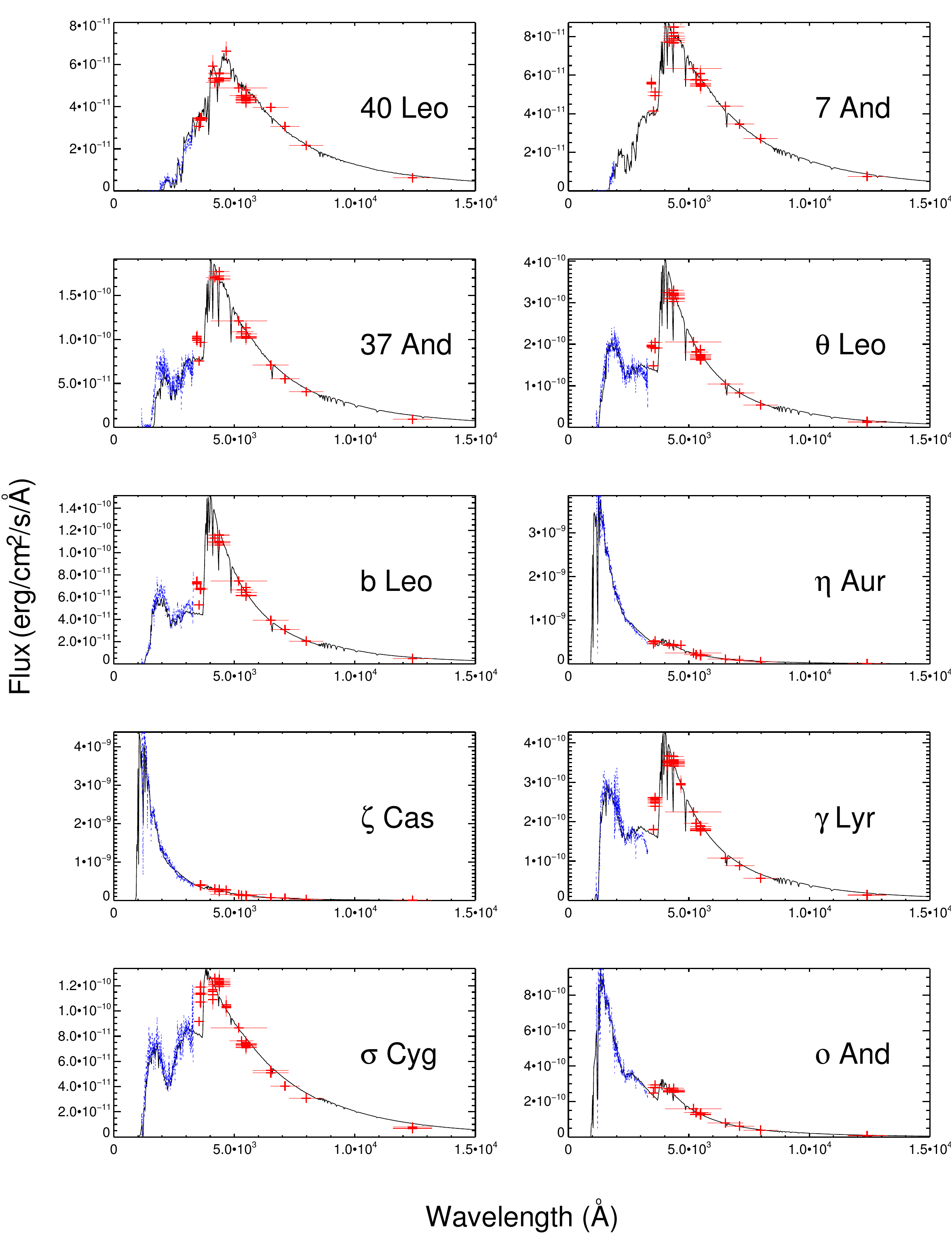}
  \caption{SED fits for the stars in our sample. Continuous line represent the best model fit to the  optical and infrared broadband photometric data (red crosses). Horizontal error bars stand for the effective width of the corresponding broadband filter. The ultraviolet co-added low resolution IUE spectra are plotted as blue dots.}
  \label{fig:sedplots}
\end{figure*}


Figure~\ref{fig:sedplots} display plots of the  resulting SED fits for the stars in the sample studied. For all the stars, the overall agreement between observed and theoretical fluxes is excellent, both in the ultraviolet and visible/near-infared regions. The best fit physical parameters $T_\textrm{eff}^{model}$ and $\log\,g$ are also in excellent agreement with spectroscopic determinations, with the only exception of $\gamma$\,Lyr, that shows a $\log\,g$ value smaller than the one derived from spectroscopy.  Temperature measurements of $\gamma$\,Lyr are widely spread in the range 9346\,K-12715\,K \citep{Ammons2006,Koleva2012}, with associated $\log$\,g values in the interval $3.5-4.11$, that appear abnormally high for a giant star. Despite the SED fit method followed should not be considered accurate enough in terms of $\log\,g$ estimation, the large discrepancy observed seems compatible with a lower expected surface gravity for a giant star than the spectroscopic values (see Table~\ref{tab:params}). The atmospheric parameters of this object will be discussed again in Section~\ref{ssec:massage}.

\begin{table*}
\caption{Estimated bolometric fluxes from SED fits of ATLAS9 grids of stellar atmosphere models computed by \citet{Castelli2003}, using optical and near-infrared photometry ($F_{bol}^{SED}$) and low-resolution \textit{IUE} ultraviolet spectra in \textit{short} (1200-1900{\AA}) and \textit{long} (1900-3300{\AA}). The table includes the reddening used for each star, as well as the best fit $T_\mathrm{eff}^{model}$ and $\log\,g$ parameter values. }
\label{tab:bolflux}
\begin{tabular}{l r@{$\pm$}l c c r@{$\pm$}l r@{$\pm$}l r@{$\pm$}l }
\hline
\hline
\textbf{Star}    & \multicolumn{2}{c}{\textbf{E(B-V)}} & \multicolumn{2}{c}{\textbf{Spectrophotometry}} & \multicolumn{4}{c}{\textbf{Model Fit}}  & \multicolumn{2}{c}{\textbf{$F_{bol}$}} \\
\textbf{Name} &  \multicolumn{2}{c}{\scriptsize \textbf{(mag)}} & \textbf{$N_{phot}$} & \textbf{$N_{UV}^{short}$} /\textbf{$N_{UV}^{long}$} &  \multicolumn{2}{c}{T$_{model}$ (K)} & \multicolumn{2}{c}{$\log\,g$ (c.g.s.)} & \multicolumn{2}{c}{\tiny \textbf{(10$^{\text{-}8}$erg\,cm$^{\text{-}2}$\,s$^{\text{-}1}$)}} \\
\hline
40\,Leo & 0.004 & 0.001 & 82 & 1/1 & 5812 & 126 & 4.29 & 0.29 & 30.9 & 0.9 \\
7\,And & 0.006 & 0.001 & 67 & 3/- & 7024 & 354 & 4.00 & 0.17 & 41 & 2 \\
37\,And & 0.004 & 0.001 & 43 & 1/1 & 7577 & 291 & 4.00 & 1.06 & 80 & 5 \\
$\theta$\,Leo & 0.005 & 0.001 & 65 & 2/1 & 9147 & 47 & 3.55 & 0.27 & 147 & 4 \\
b\,Leo & 0.005 & 0.001 & 50 & 3/4 & 8759 & 37 & 3.83 & 0.35 & 51.0 & 1.2 \\
$\eta$\,Aur & 0.014 & 0.001 & 85 & 11/40 & 16212 & 63 & 4.40 & 0.30 & 831 & 11 \\
$\zeta$\,Cas & 0.04 & 0.04 & 54 & 1/4 & 18562 & 151 & 4.02 & 0.28 & 693 & 20 \\
\hline
 $\gamma$\,Lyr & 0.017 & 0.004 & 83 & 1/1 & 9640 & 60 & 2.64 & 0.35 & 215 & 4 \\
 $\sigma$\,Cyg  & 0.13 & 0.05 & 78 & 1/1 & 9990 & 0 & 1.90 & 0.06 & 133 & 3 \\
$o$\,And & 0.049 & 0.022 & 45 & 13/12 & 13812 & 45 & 3.17 & 0.10 & 383 & 5 \\
\hline
\end{tabular}
\end{table*}


\subsection{Luminosities, temperatures and linear radii}

The combination of the derived values for stellar angular diameters $\theta_{\mathrm{LD}}$ and bolometric fluxes $F_{bol}$ with the \textit{Hipparcos} parallaxes $\pi$ provides us with estimations of linear radii, luminosities and effective temperatures for the stars in our sample. Measured angular limb-darkened diameters $\theta_{LD}$ are transformed into linear radius $R$ for each star using the \textit{Hipparcos} \citep{vanLeeuwen2007} parallaxes. The absolute luminosity is computed from the known distance $d$ to the object and the bolometric flux $F_{bol}$ by solving

\begin{equation}\label{eq:luminosity}
L=4\pi d^2 F_{bol}
\end{equation}

Finally, the combination of angular diameter $\theta_{\mathrm{LD}}$ with the estimated bolometric flux $F_{bol}$ allows us to measure the effective temperature of a star, defined according to

\begin{equation}\label{eq:teff}
T_{eff}=\left(4\frac{F_{bol}}{\sigma_B\theta_{LD}^2}\right)^{1/4}
\end{equation}

\begin{figure}
  \centering  
  \includegraphics[width=\columnwidth,angle=0]{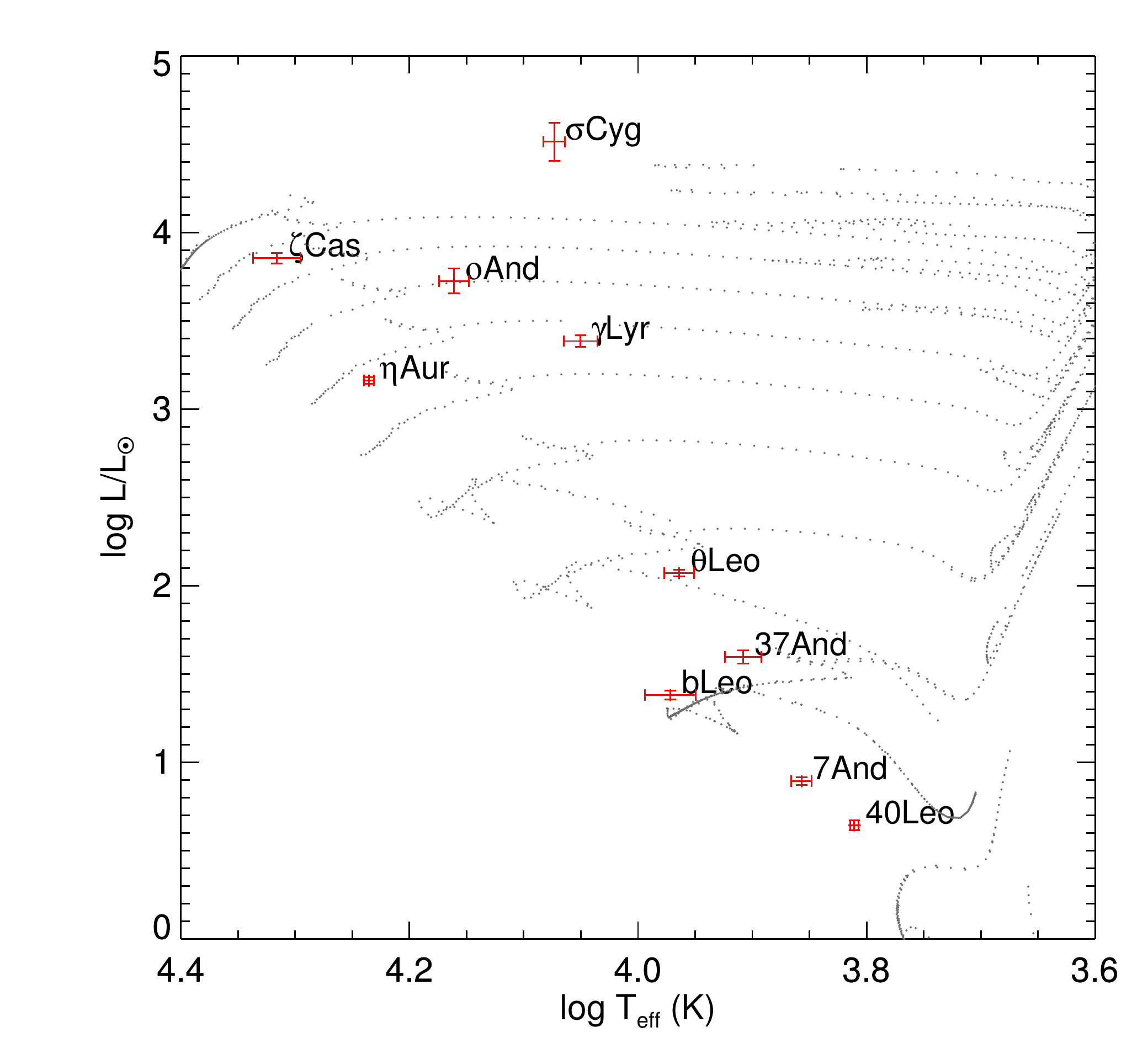}
  \caption{Hertzprung-Russell diagram containing luminosity and effective temperature derived for the stars in our study (see Table~\ref{tab:radlumteff}). Solar-metallicity PARSEC \citep{Bressan2012} evolutionary tracks for masses in the range 1-10\,$M_\odot$ in steps of 1\,$M_\odot$ are shown as grey dotted lines.}
  \label{fig:lumteff}
\end{figure}

The values obtained are summarised in Table~\ref{tab:radlumteff}. Figure~\ref{fig:lumteff} display the results in a Hertzprung-Russell diagram.

\begin{table}
\caption{Linear radii, luminosities and effective temperatures.}
\label{tab:radlumteff}
\centering
\begin{tabular}{l  r@{$\pm$}l r@{$\pm$}l r@{$\pm$}l }
\hline
\hline
\textbf{Star} & \multicolumn{2}{c}{$R\,(R_\odot)$} & \multicolumn{2}{c}{$L\,(L_\odot)$} & \multicolumn{2}{c}{$T_\mathrm{eff}\,(K)$} \\
\hline
40\,Leo                     	& 1.68 & 0.07 & 4.4 & 0.9    & 6450 & 140 \\
7\,And                       	& 1.71 & 0.02 & 7.8 & 0.6    & 7380 & 90 \\
37\,And          		& 3.03 & 0.11 & 40 & 3         & 8320 & 150 \\
$\theta$\,Leo       	& 4.03 & 0.10 & 118 & 5       & 9480 & 120 \\
$b$\,Leo                  	& 1.80 & 0.07 & 24.1 & 1.4 & 9540 & 180 \\
$\eta$\,Aur            	& 3.64 & 0.10 & 1450 & 70 & 18660 & 230 \\
$\zeta$\,Cas       	& 6.1 & 0.3 & 7200 & 900    & 21500 & 400 \\
\hline
$\gamma$\,Lyr   	& 15.4 & 0.8 & 2430 & 190 & 10330 & 80 \\
$\sigma$\,Cyg     	& 50 & 9 & 33000 & 8000    & 10940 & 180\\
$o$\,And                     & 11.5 & 1.3 & 5300 & 900 & 14540 & 170 \\
\hline
\end{tabular}
\end{table}


\subsection{Effective temperature}\label{ssec:teff}

\begin{figure}
  \centering  
  \includegraphics[width=\columnwidth,angle=0]{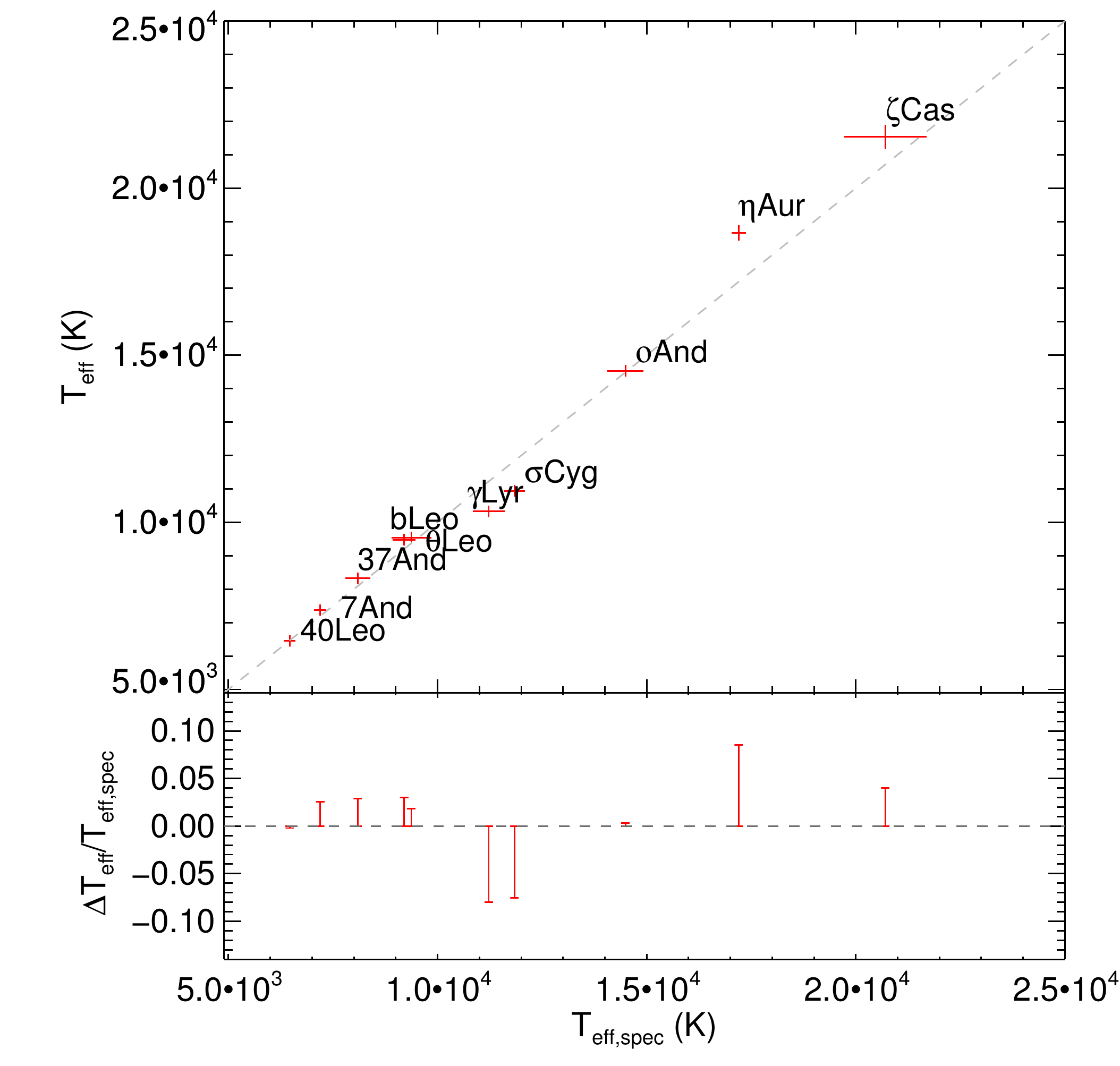}
  \caption{(\textit{Top}) Comparison of effective temperature estimations derived in this paper with respect to spectroscopic determinations (see Table~\ref{tab:params}). (\textit{Bottom}) Relative difference between both estimations of effective temperature.}
  \label{fig:compteff}
\end{figure}

Figure~\ref{fig:compteff} displays a comparison of the derived effective temperature values derived from our measurements of  $\theta_{\mathrm{LD}}$ and $F_{bol}$ and values determined using spectroscopy listed in Table~\ref{tab:params}. There is an excellent agreement between both sets of measurements, specially for stars of spectral types A0 or later. The average and median deviation are $\simeq0.7\%$ and $\simeq 2.1\%$ respectively, with a scatter of $5\%$. For main-sequence stars, the scatter reduces to $\simeq2.5\%$.

Given the effective temperatures in excess of 10000\,K of stars with spectral types earlier than A0, a  significant fraction of the total emergent flux is radiated in the ultraviolet region, where systematics caused by limitations of plane-parallel stellar atmosphere models are conspicuous, especially in the case of giants or supergiants. The physical parameters describing the model stellar atmosphere best fit do not necessarily represent the best description of the actual physical parameters of the star. Nonetheless, the $F_{bol}^{1/4}$ dependency of $T_\mathrm{eff}$ makes  the determination of effective temperature to remain robust in spite of the simplified assumptions contained in the stellar atmosphere models used.

Table~\ref{tab:zorec} compares $T_\mathrm{eff}$ derived in our study with results presented in \citet{Zorec2009} for a subset of the 5 overlapping objects in both samples, with $T_\mathrm{eff}>$9000\,K. The agreement of both sets of measurements is excellent. Small disagreements are likely due to the simultaneous estimation of the photosphere diameter $\theta^f$ and the identification of model temperature $T_\mathrm{eff}^{model}$ and effective temperature $T_\mathrm{eff}$ made in \citet{Zorec2009}. In our case, including a direct measurement of the limb-darkened diameter, assumed to represent the photosphere diameter and estimating $T_\mathrm{eff}$ exclusively from the estimation of the total amount of radiated energy reduce considerably the model dependence of our results, that is limited to the use of theoretical spectra to fit the observed fluxes. 
 
\begin{table}
\caption{Comparison of derived effective temperatures with results presented in \citet{Zorec2009}.}
\label{tab:zorec}
\centering
\begin{tabular}{l r@{$\pm$}l r@{$\pm$}l}
\hline
\hline
\textbf{Star} & \multicolumn{2}{c}{$T_{eff}\,(K)$} & \multicolumn{2}{c}{$T_{eff}^{Zorec}\,(K)$} \\
\hline
$\theta$\,Leo       	& 9480 & 120 & 9180 & 290\\
$\eta$\,Aur            	& 18660 & 230 & 17940 & 1070 \\
$\zeta$\,Cas       	& 21500 & 400 & 21850 & 1390 \\
\hline
$\gamma$\,Lyr   	& 10330 & 80 & 10000 & 350\\
$\sigma$\,Cyg     	& 10940 & 180 & 11170 & 450\\
\hline
\end{tabular}
\end{table}

\subsection{Color-effective temperature relations}\label{ssec:colorteff}

\begin{figure}
  \centering  
   \includegraphics[width=\columnwidth]{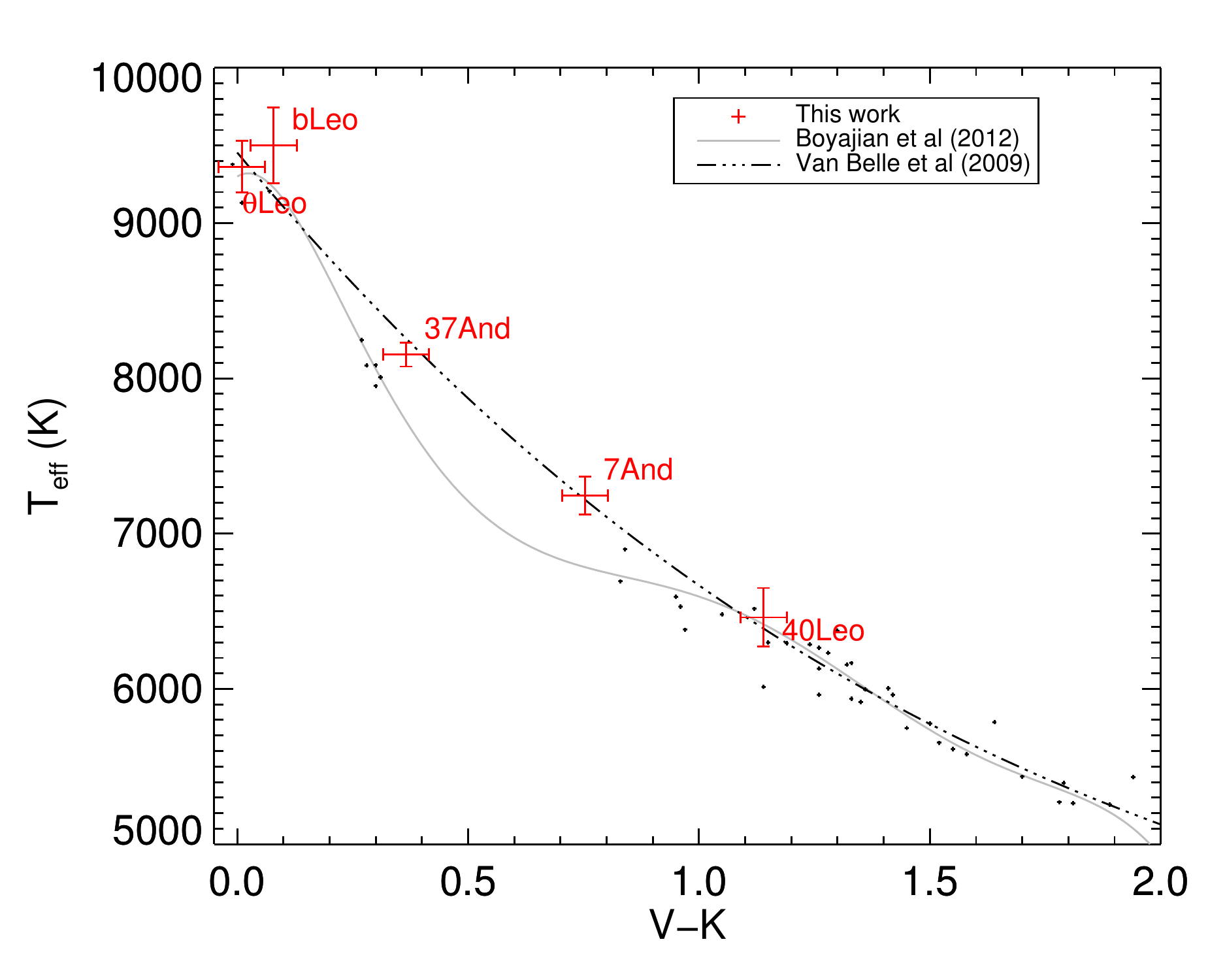}
  \caption{Plot of the effective temperature versus $V-K$ color for the main sequence stars in the studied sample with $T_\mathrm{eff}$ below 10000\,K. The solid line represents the empirical relation found in \citet{Boyajian2012a} by means of a sixth-order polynomial fit to a sample of 44 A- to G-type stars (small black dots). Triple dot-dash line represents the third-order polynomial fit presented in \citet{vanBelle2009}.}
  \label{fig:teff_vk}
\end{figure}

\begin{figure}
  \centering  
  \includegraphics[width=\columnwidth]{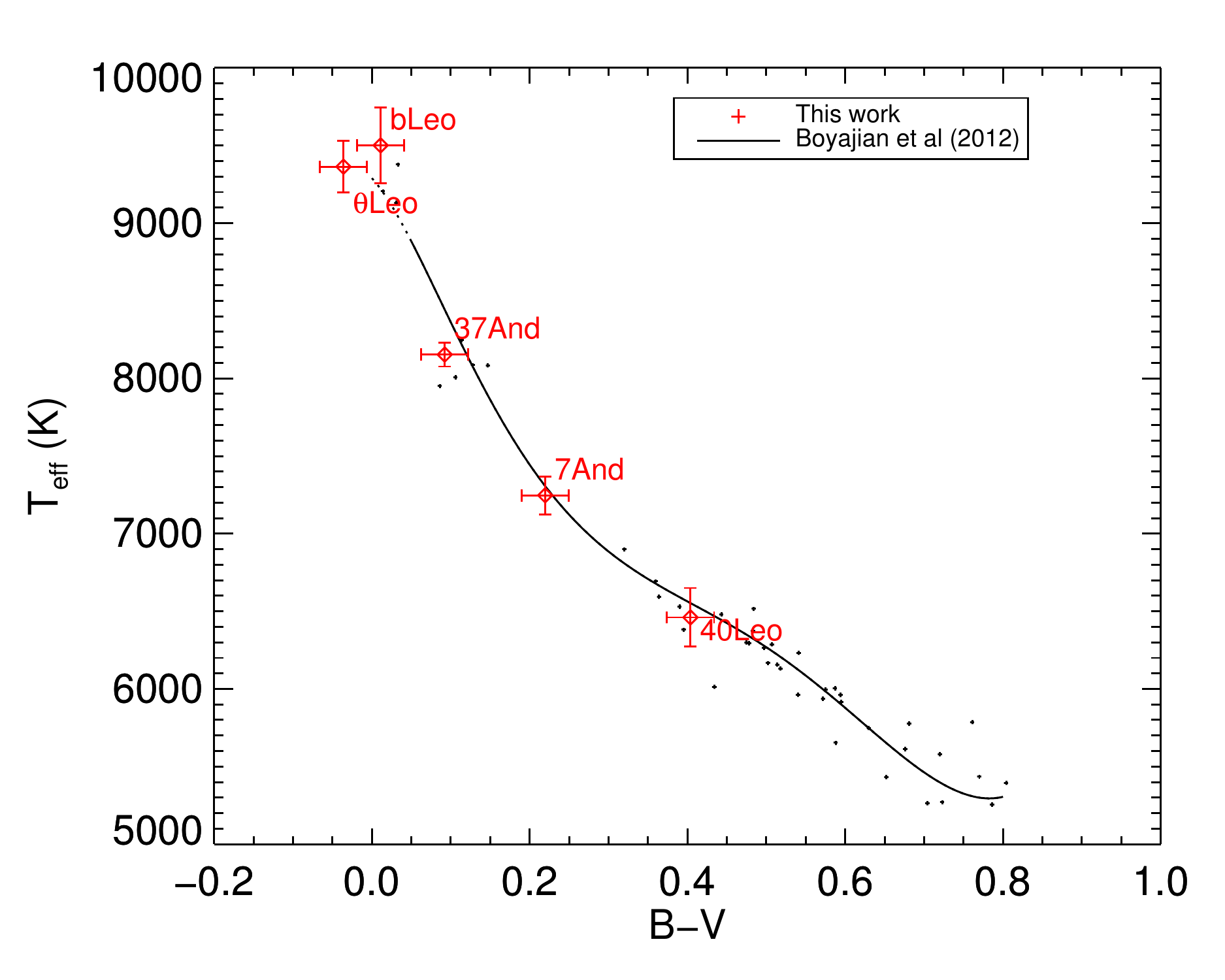}
  \caption{Effective temperature versus $B-V$ color for the main sequence stars in the observed sample with $T_\mathrm{eff}$ below 10000\,K. Data from \citet{Boyajian2012a} (small black dots), as well as empirical relation based on a sixth-order polynomial fit (solid line) are also included.}
  \label{fig:teff_bv}
\end{figure}

Whereas the empirical temperature scale for giant stars seems to be firmly established with uncertainties under 2.5\% \citep{Code1976,Underhill1979,vanBelle1999a}, the same cannot be said for main sequence stars (luminosity classes IV-V). For those stars in our sample with $T_\mathrm{eff}<$10000\,K (spectral types A0 or later), we have contrasted the computed effective temperatures with empirical colour-temperature relations presented in \citet{Boyajian2012a} and \citet{vanBelle2009} using ($V-K$) and ($B-V$) colours. Figures~\ref{fig:teff_vk} and \ref{fig:teff_bv} display effective temperature versus ($V-K$) and ($B-V$)  colours for the stars in our sample lying in the temperature range considered, as well as the results presented in \citet{Boyajian2012a}, fitted to a sixth-order polynomial in ($V-K$) and ($B-V$) respectively. Figure~\ref{fig:teff_vk} adds the empirical relation found by \citet{vanBelle2009}, using a third-order polynomial. Most of the stars in both previous studies are cooler than 7000\,K, and both empirical relations nearly overlap in that range of temperatures. The same does not apply for earlier spectral types, as the two fits differ quite significantly between 6500\,K and 8500\,K, where there were no stars in their samples. Three of the stars in our study (37\,And, 7\,And and 40\,Leo) have temperatures within this range. Their location in the $T_\mathrm{eff}$-($V-K$) diagram (Figure~\ref{fig:teff_vk}) shows strong agreement with the more conservative third-order polynomial relation presented by \citet{vanBelle2009}. On the other hand, our data is fully consistent with the $T_\mathrm{eff}$-($B-V$) sixth-order polynomial relation of \citet{Boyajian2012a} (Figure~\ref{fig:teff_bv}). Observational effects induced by ubiquitous fast rotation among stars  earlier than F6 could explain the disagreement with \citet{Boyajian2012a} sixth-order ($V-K$) polynomial relation. Most of the A-type stars studied in that paper show high projected rotational velocities ($v\sin i$), compatible with fast rotation seen at a lower inclination angle (closer to be equator-on) with respect to the same spectral type stars contained in our sample. Further accurate estimations of $T_\mathrm{eff}$ for a larger sample of main sequence stars earlier than F5 will significantly improve the empirical temperature scale for these stars.

\subsection{Masses and ages}\label{ssec:massage}

\begin{figure}
  \centering  
  \includegraphics[height=\columnwidth,angle=270]{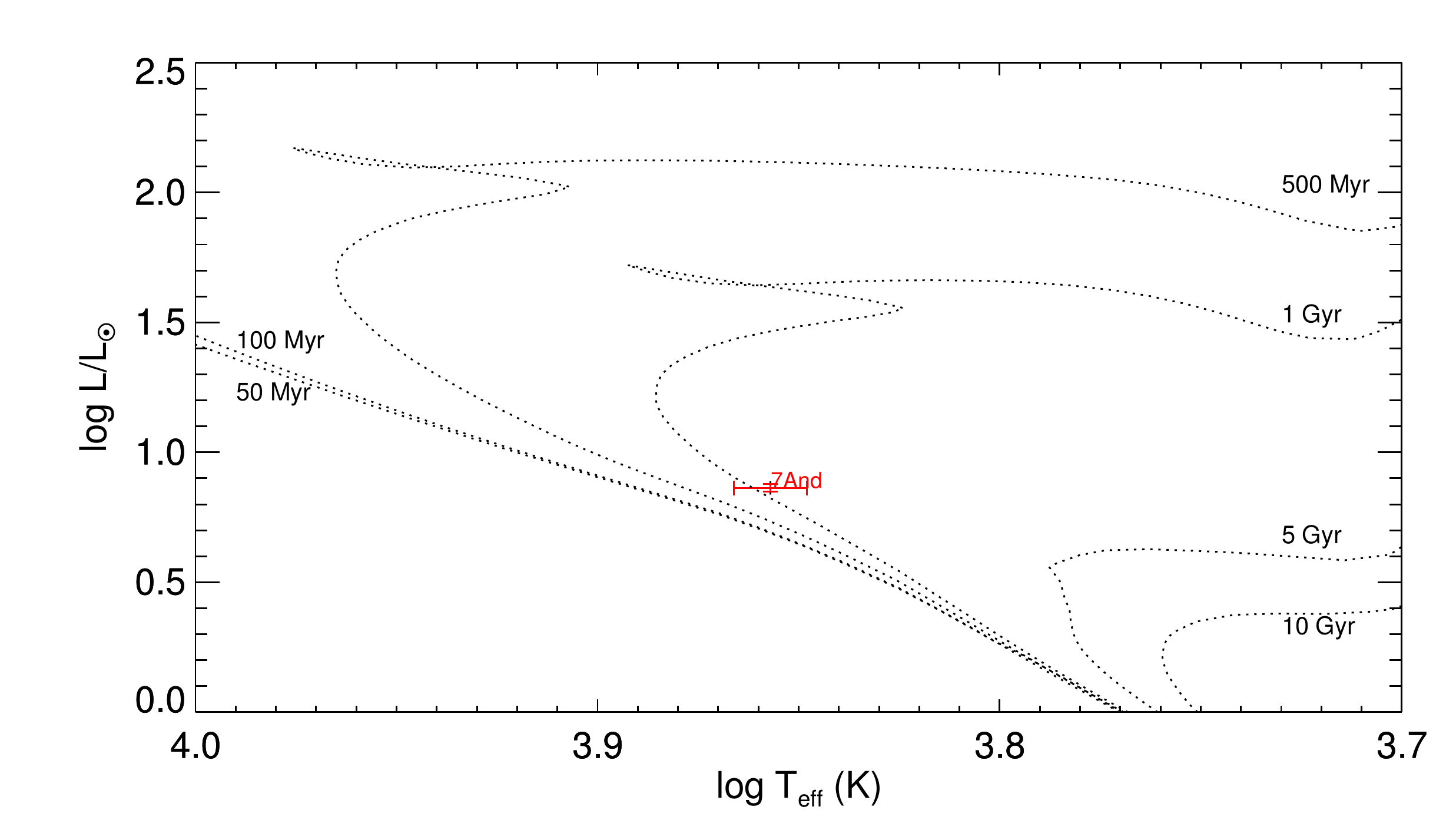}
  \caption{Luminosity-temperature diagram showing the isochrones generated for 7\,And.}
  \label{fig:massage_7and}
\end{figure}

We have estimated the  stellar masses ($M_{\star,iso}$) and ages for our target sample by fitting the inferred luminosity and effective temperature values to PARSEC\footnote{\tt{http://stev.oapd.inaf.it/cmd}} isochrones \citep{Bressan2012}. Figure~\ref{fig:massage_7and} illustrates the method using 7\,And as an example. For each object in our sample, we have computed a grid of isochrones equally spaced in $\log age$ ($age$ in years) in the range 1\,Myr-10\,Gyr, and adopt the best fit to the luminosity and effective temperature computed values. We have assumed solar metallicity for all the stars in the sample. Uncertainties are derived by considering solutions at 1-$\sigma$ separation in both luminosity and temperature dimensions. Results obtained are displayed in Table~\ref{tab:massages}. 

\begin{table}\caption{Isochrone masses and ages.}
\label{tab:massages}
\begin{center}
\begin{tabular}{l  r@{$\pm$}l r@{$\pm$}l  }
\hline
\hline
\textbf{Star} & \multicolumn{2}{c}{$M\,(M_\odot)$} & \multicolumn{2}{c}{$Age\,(Myr)$}  \\
\hline
40\,Leo                     	& 1.35 & 0.06 & 2630 & 210 \\
7\,And                       	& 1.6 & 0.1 & 1120 & 30 \\
37\,And          		& 2.21 & 0.09 & 724 & 21 \\
$\theta$\,Leo       	& 2.8 & 0.1 & 407 & 12 \\
$b$\,Leo                 	& 2.11 & 0.06 & 195 & 15 \\
$\eta$\,Aur            	& 5.6 & 0.1 & 41 & 6 \\
$\zeta$\,Cas       	& 8.96 & 0.13 & 22.9 & 1.2 \\
\hline
$\gamma$\,Lyr   	& 5.76 & 0.13 & 74.8 & 5.1 \\
$\sigma$\,Cyg     	& 11.2 & 0.2 & 19.1 & 0.6 \\
$o$\,And                 	& 6.5 & 0.5 & 52 & 9 \\
\hline
\end{tabular}
\end{center}
\end{table}

\begin{figure}
  \centering 
  \includegraphics[width=\columnwidth]{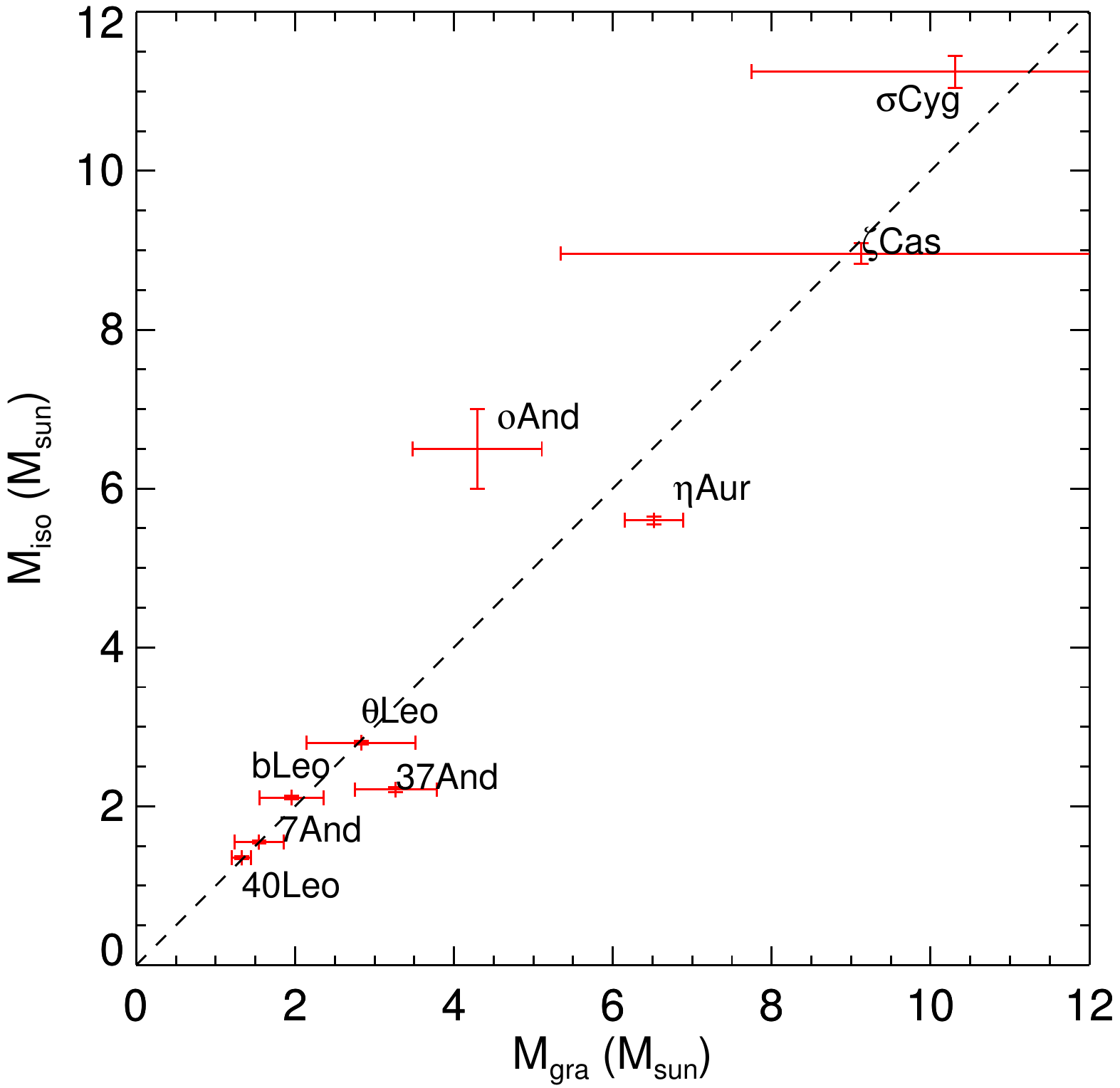}
  \caption{Comparison of mass estimations using isochrone fit ($M_{iso}$) and the gravitational mass $M_{gra}$, computed using spectroscopic surface gravity and linear stellar radii (Equation~\ref{eq:gravmass}). The dashed line represents identity. $\gamma$\,Lyr ($M_{gra}=$55\,$M_\odot$, $M_{iso}=$5.76$\pm$0.13\,$M_\odot$) is not shown in the plot.}
  \label{fig:miso_mgra}
\end{figure}

Using the linear radii estimated from combined measurements of angular diameter and parallax and the spectrocospic determination of $\log g$, we can estimate the so-called \textit{gravity mass} \citep{vanBelle2007} of each object according to

\begin{equation}\label{eq:gravmass}
g=G\frac{M_{gra}}{R_\star^2}
\end{equation}

\noindent  where $G$ is the gravitational constant, and $M_{gra}$ and $R_\star$ stand for the mass and linear radius of the star considered. Figure~\ref{fig:miso_mgra} displays the relation between isochrone and gravitational masses. Comparison of masses determined by both methods show a remarkable good agreement for all the objects in the sample except $\gamma$\,Lyr. The reason for the large disagreement in that case is most likely related to the spectroscopic determination of $\log\,g$, since values of $3.5$, $4.11$ and $3.68$ reported by different authors \citep[][ respectively]{Balachandran1986,Prugniel2011,Koleva2012} are clearly discrepant with the values expected for a B9 giant, which combined with our linear radius estimation result in unphysical gravity mass values of $M_{gra}=$27\,$M_\odot$, 41\,$M_\odot$ and 111\,$M_\odot$ (55\,$M_\odot$ for the mean value $\log g$=3.81). However, the value of $\log\,g$ derived from our SED fit yields to a gravity mass value of $\simeq$4.4\,$M_\odot$, in better agreement with the $M_{iso}$ value obtained.

\section{Conclusions}
\label{sec:conclusions}

We have presented results of a pilot study on a sample of 10 stars (7 main sequence or sub giants and 3 supergiants or giants) with spectral types between B2 and F6. For all the objects in the sample, we have measured submilliarcsecond angular diameters and bolometric fluxes with an average precision of 2.3\% and 2.1\% respectively. Combined with \textit{Hipparcos} parallaxes, we have derived fundamental stellar parameters, linear radii, effective temperatures, and luminosities with 8\%, 3\% and 5\% average relative uncertainties. Finally, we have fitted PARSEC isochrones to the values of temperature and luminosity found in order to obtain estimates of mass and age for every star in the sample. Our findings can be summarised as follows:

\begin{enumerate}
\item Measured diameters show generally good agreement with predictions based on colour relations or SED fits, as well as with previous interferometric measurements. $V^2$ data points beyond the first null for $\theta$\,Leo enable simultaneous estimation of diameter and linear limb-darkening coefficient $\mu$. Whereas the derived diameter is less than 1\% larger than the fixed limb darkening estimation, the estimated value for $\mu$ exceeds by $\simeq$20\% the value interpolated in \citet{Claret2011} using stellar atmosphere parameters. This enhanced limb darkening, together with the star's low projected rotational velocity, suggests that $\theta$\,Leo is in fact a fast rotating star viewed nearly pole-on.
\item We derived bolometric fluxes from SED fits of theoretical ATLAS9 stellar atmosphere models to optical and near-infrared photometric data, as well as including flux calibrated \textit{IUE} ultraviolet spectra, accounting for the effects of interstellar reddening. The mean absolute deviation between derived effective temperatures and estimates using spectroscopy  is $\simeq$3\%. The agreement is excellent for main sequence stars, and only for two of the B giants, $\gamma$\,Lyr, and $\sigma$\,Cyg and the main-squence star $\eta$\,Aur, the discrepancy is as large as $\simeq$7\%. Comparison with \citet{Zorec2009}, for the subset of overlapping objects from our sample reveals an excellent agreement in the results of both studies, despite some subtle differences in the modelling assumptions.
\item  Our derived effective temperatures for A and F type stars confirm the sixth-order polynomial ($B-V$) color-temperature relation presented in \citet{Boyajian2012a}. However, there is a clear disagreement with their sixth-order ($V-K$)-temperature relation, and our data clearly favours the ($V-K$) cubic polynomial presented by \citet{vanBelle2009}. Different inclination angles of the polar axis between Boyajian's sample of A and early F stars with respect to ours, together with the effects of fast rotation in these stars might explain the observed discrepancy. New observations of main-sequence stars in this range of temperature, correcting the scarcity of available data should provide the basis for a more detailed empirical colour-temperature relation.
\item The PARSEC model isochrone fit in the temperature-luminosity plane provides mass values that are consistent with the estimated gravitational masses computed from spectroscopically determined $\log g$, showing larger discrepancies for larger mass values. The unphysical result obtained for $\gamma$\,Lyr gravitational mass, together with the large scatter in temperature estimations using spectroscopy questions the accuracy of $\gamma$\,Lyr atmosphere parameters, and suggests strong degeneracy between $\log g$ and $T_\mathrm{eff}$. Additionally, a $\log,g$ estimation based on the SED fit to spectrophotometric data seems to give a gravity mass that is more consistent with the value obtained by means of the isochrone fitting.
\end{enumerate}


\section*{Acknowledgements}

This research has made use of the SIMBAD database and the VizieR catalogue access tool, operated at CDS, Strasbourg, France.
Some of the data presented in this paper were obtained from the Multimission Archive at the Space Telescope Science Institute (MAST). STScI is operated by the Association of Universities for Research in Astronomy, Inc., under NASA contract NAS5-26555. Support for MAST for non-HST data is provided by the NASA Office of Space Science via grant NAG5-7584 and by other grants and contracts. This publication makes use of data products from the Two Micron All Sky Survey, which is a joint project of the University of Massachusetts and the Infrared Processing and Analysis Center/California Institute of Technology, funded by the National Aeronautics and Space Administration and the National Science Foundation.
The CHARA Array is funded by the National Science Foundation through NSF grant AST-0606958, by Georgia State University through the College of Arts and Sciences, and by the W.M. Keck Foundation. We acknowledge the support of the Australian Research Council.
V.M. is supported by an International Denison Postgraduate Award.





\begin{thebibliography}{}

\bibitem[\protect\citeauthoryear{{Adelman}}{{Adelman}}{1986}]{Adelman1986}
{Adelman} S.~J.,  1986, \aaps, 64, 173

\bibitem[\protect\citeauthoryear{{Adelman}}{{Adelman}}{1988}]{Adelman1988}
{Adelman} S.~J.,  1988, \mnras, 230, 671

\bibitem[\protect\citeauthoryear{{Adelman}, {Pintado}, {Nieva}, {Rayle} \&
  {Sanders} Jr.}{{Adelman} et~al.}{2002}]{Adelman2002}
{Adelman} S.~J.,  {Pintado} O.~I.,  {Nieva} M.~F.,  {Rayle} K.~E.,    {Sanders}
  Jr. S.~E.,  2002, \aap, 392, 1031

\bibitem[\protect\citeauthoryear{{Allende Prieto} \& {Lambert}}{{Allende
  Prieto} \& {Lambert}}{1999}]{AllendePrieto1999}
{Allende Prieto} C.,  {Lambert} D.~L.,  1999, \aap, 352, 555

\bibitem[\protect\citeauthoryear{{Ammons}, {Robinson}, {Strader}, {Laughlin},
  {Fischer} \& {Wolf}}{{Ammons} et~al.}{2006}]{Ammons2006}
{Ammons} S.~M.,  {Robinson} S.~E.,  {Strader} J.,  {Laughlin} G.,  {Fischer}
  D.,    {Wolf} A.,  2006, \apj, 638, 1004

\bibitem[\protect\citeauthoryear{{Aufdenberg} et~al.,}{{Aufdenberg}
  et~al.}{2006}]{Aufdenberg2006}
{Aufdenberg} J.~P.  et~al., 2006, \apj, 645, 664

\bibitem[\protect\citeauthoryear{{Balachandran}, {Lambert}, {Tomkin} \&
  {Parthasarathy}}{{Balachandran} et~al.}{1986}]{Balachandran1986}
{Balachandran} S.,  {Lambert} D.~L.,  {Tomkin} J.,    {Parthasarathy} M.,
  1986, \mnras, 219, 479

\bibitem[\protect\citeauthoryear{{Balona} \& {Dziembowski}}{{Balona} \&
  {Dziembowski}}{1999}]{Balona1999}
{Balona} L.~A.,  {Dziembowski} W.~A.,  1999, \mnras, 309, 221

\bibitem[\protect\citeauthoryear{{Bazot} et~al.,}{{Bazot}
  et~al.}{2011}]{Bazot2011}
{Bazot} M.  et~al., 2011, \aap, 526, L4

\bibitem[\protect\citeauthoryear{{Berger} et~al.,}{{Berger}
  et~al.}{2006}]{Berger2006}
{Berger} D.~H.  et~al., 2006, \apj, 644, 475

\bibitem[\protect\citeauthoryear{{Bessell} \& {Murphy}}{{Bessell} \&
  {Murphy}}{2012}]{Bessell2012}
{Bessell} M.,  {Murphy} S.,  2012, \pasp, 124, 140

\bibitem[\protect\citeauthoryear{{Bessell}, {Castelli} \& {Plez}}{{Bessell}
  et~al.}{1998}]{Bessell1998}
{Bessell} M.~S.,  {Castelli} F.,    {Plez} B.,  1998, \aap, 333, 231

\bibitem[\protect\citeauthoryear{{Blackwell} \& {Lynas-Gray}}{{Blackwell} \&
  {Lynas-Gray}}{1998}]{Blackwell1998}
{Blackwell} D.~E.,  {Lynas-Gray} A.~E.,  1998, \aaps, 129, 505

\bibitem[\protect\citeauthoryear{{Boden}}{{Boden}}{2003}]{Boden2003}
{Boden} A.~F.,  2003. p.~151

\bibitem[\protect\citeauthoryear{{Boyajian} et~al.,}{{Boyajian}
  et~al.}{2012a}]{Boyajian2012a}
{Boyajian} T.~S.  et~al., 2012a, \apj, 746, 101

\bibitem[\protect\citeauthoryear{{Boyajian} et~al.,}{{Boyajian}
  et~al.}{2012b}]{Boyajian2012b}
{Boyajian} T.~S.  et~al., 2012b, \apj, 757, 112

\bibitem[\protect\citeauthoryear{{Bressan}, {Marigo}, {Girardi}, {Salasnich},
  {Dal Cero}, {Rubele} \& {Nanni}}{{Bressan} et~al.}{2012}]{Bressan2012}
{Bressan} A.,  {Marigo} P.,  {Girardi} L.,  {Salasnich} B.,  {Dal Cero} C.,
  {Rubele} S.,    {Nanni} A.,  2012, \mnras, 427, 127

\bibitem[\protect\citeauthoryear{{Castelli} \& {Kurucz}}{{Castelli} \&
  {Kurucz}}{2003}]{Castelli2003}
{Castelli} F.,  {Kurucz} R.~L.,  2003, in {Piskunov} N.,  {Weiss} W.~W.,
  {Gray} D.~F.,  eds,  IAU Symposium Vol. 210, Modelling of Stellar
  Atmospheres. p.~20P

\bibitem[\protect\citeauthoryear{{Che} et~al.,}{{Che} et~al.}{2011}]{Che2011}
{Che} X.  et~al., 2011, \apj, 732, 68

\bibitem[\protect\citeauthoryear{{Claret} \& {Bloemen}}{{Claret} \&
  {Bloemen}}{2011}]{Claret2011}
{Claret} A.,  {Bloemen} S.,  2011, \aap, 529, A75

\bibitem[\protect\citeauthoryear{{Clark}, {Tarasov} \& {Panko}}{{Clark}
  et~al.}{2003}]{Clark2003}
{Clark} J.~S.,  {Tarasov} A.~E.,    {Panko} E.~A.,  2003, \aap, 403, 239

\bibitem[\protect\citeauthoryear{{Code}, {Bless}, {Davis} \& {Brown}}{{Code}
  et~al.}{1976}]{Code1976}
{Code} A.~D.,  {Bless} R.~C.,  {Davis} J.,    {Brown} R.~H.,  1976, \apj, 203,
  417

\bibitem[\protect\citeauthoryear{{Cohen}, {Wheaton} \& {Megeath}}{{Cohen}
  et~al.}{2003}]{Cohen2003}
{Cohen} M.,  {Wheaton} W.~A.,    {Megeath} S.~T.,  2003, \aj, 126, 1090

\bibitem[\protect\citeauthoryear{{Cottrell} \& {Sneden}}{{Cottrell} \&
  {Sneden}}{1986}]{Cottrell1986}
{Cottrell} P.~L.,  {Sneden} C.,  1986, \aap, 161, 314

\bibitem[\protect\citeauthoryear{{Cruzal{\`e}bes}, {Jorissen}, {Sacuto} \&
  {Bonneau}}{{Cruzal{\`e}bes} et~al.}{2010}]{Cruzalebes2010}
{Cruzal{\`e}bes} P.,  {Jorissen} A.,  {Sacuto} S.,    {Bonneau} D.,  2010,
  \aap, 515, A6

\bibitem[\protect\citeauthoryear{{Davis}}{{Davis}}{1997}]{Davis1997}
{Davis} J.,  1997, in {Bedding} T.~R.,  {Booth} A.~J.,   {Davis} J.,  eds,
  Vol. 189, IAU Symposium. pp 31--38

\bibitem[\protect\citeauthoryear{{Derekas} et~al.,}{{Derekas}
  et~al.}{2011}]{Derekas2011}
{Derekas} A.  et~al., 2011, Science, 332, 216

\bibitem[\protect\citeauthoryear{{Domiciano de Souza}, {Vakili}, {Jankov},
  {Janot-Pacheco} \& {Abe}}{{Domiciano de Souza}
  et~al.}{2002}]{DomicianoDeSouza2002}
{Domiciano de Souza} A.,  {Vakili} F.,  {Jankov} S.,  {Janot-Pacheco} E.,
  {Abe} L.,  2002, \aap, 393, 345

\bibitem[\protect\citeauthoryear{{Drimmel}, {Cabrera-Lavers} \&
  {L{\'o}pez-Corredoira}}{{Drimmel} et~al.}{2003}]{Drimmel2003}
{Drimmel} R.,  {Cabrera-Lavers} A.,    {L{\'o}pez-Corredoira} M.,  2003, \aap,
  409, 205

\bibitem[\protect\citeauthoryear{{Eggleton} \& {Tokovinin}}{{Eggleton} \&
  {Tokovinin}}{2008}]{Eggleton2008}
{Eggleton} P.~P.,  {Tokovinin} A.~A.,  2008, \mnras, 389, 869

\bibitem[\protect\citeauthoryear{{Erspamer} \& {North}}{{Erspamer} \&
  {North}}{2003}]{Erspamer2003}
{Erspamer} D.,  {North} P.,  2003, \aap, 398, 1121

\bibitem[\protect\citeauthoryear{{Fitzpatrick}}{{Fitzpatrick}}{1999}]{Fitzpatrick1999}
{Fitzpatrick} E.~L.,  1999, \pasp, 111, 63

\bibitem[\protect\citeauthoryear{{Fitzpatrick} \& {Massa}}{{Fitzpatrick} \&
  {Massa}}{2005}]{Fitzpatrick2005}
{Fitzpatrick} E.~L.,  {Massa} D.,  2005, \aj, 129, 1642

\bibitem[\protect\citeauthoryear{{Fr{\'e}mat}, {Zorec}, {Hubert} \&
  {Floquet}}{{Fr{\'e}mat} et~al.}{2005}]{Fremat2005}
{Fr{\'e}mat} Y.,  {Zorec} J.,  {Hubert} A.-M.,    {Floquet} M.,  2005, \aap,
  440, 305

\bibitem[\protect\citeauthoryear{{Gardiner}, {Kupka} \& {Smalley}}{{Gardiner}
  et~al.}{1999}]{Gardiner1999}
{Gardiner} R.~B.,  {Kupka} F.,    {Smalley} B.,  1999, \aap, 347, 876

\bibitem[\protect\citeauthoryear{{Gies} \& {Lambert}}{{Gies} \&
  {Lambert}}{1992}]{Gies1992}
{Gies} D.~R.,  {Lambert} D.~L.,  1992, \apj, 387, 673

\bibitem[\protect\citeauthoryear{{G{\l}{\c e}bocki} \&
  {Gnaci{\'n}ski}}{{G{\l}{\c e}bocki} \& {Gnaci{\'n}ski}}{2005}]{Glebocki2005}
{G{\l}{\c e}bocki} R.,  {Gnaci{\'n}ski} P.,  2005, in {Favata} F.,  {Hussain}
  G.~A.~J.,   {Battrick} B.,  eds,  ESA Special Publication Vol. 560, 13th
  Cambridge Workshop on Cool Stars, Stellar Systems and the Sun. p.~571

\bibitem[\protect\citeauthoryear{{Gray}}{{Gray}}{1998}]{Gray1998}
{Gray} R.~O.,  1998, \aj, 116, 482

\bibitem[\protect\citeauthoryear{{Hanbury Brown}, {Davis} \& {Allen}}{{Hanbury
  Brown} et~al.}{1974}]{Hanbury1974a}
{Hanbury Brown} R.,  {Davis} J.,    {Allen} L.~R.,  1974, \mnras, 167, 121

\bibitem[\protect\citeauthoryear{{Hanbury Brown}, {Davis}, {Lake} \&
  {Thompson}}{{Hanbury Brown} et~al.}{1974}]{Hanbury1974b}
{Hanbury Brown} R.,  {Davis} J.,  {Lake} R.~J.~W.,    {Thompson} R.~J.,  1974,
  \mnras, 167, 475

\bibitem[\protect\citeauthoryear{{Harmanec} et~al.,}{{Harmanec}
  et~al.}{1996}]{Harmanec1996}
{Harmanec} P.  et~al., 1996, \aap, 312, 879

\bibitem[\protect\citeauthoryear{{Hill} \& {Landstreet}}{{Hill} \&
  {Landstreet}}{1993}]{Hill1993}
{Hill} G.~M.,  {Landstreet} J.~D.,  1993, \aap, 276, 142

\bibitem[\protect\citeauthoryear{{Huber} et~al.,}{{Huber}
  et~al.}{2012a}]{Huber2012b}
{Huber} D.  et~al., 2012a, \apj, 760, 32

\bibitem[\protect\citeauthoryear{{Huber} et~al.,}{{Huber}
  et~al.}{2012b}]{Huber2012a}
{Huber} D.  et~al., 2012b, \mnras, p.~L438

\bibitem[\protect\citeauthoryear{{Ireland} et~al.,}{{Ireland}
  et~al.}{2008}]{Ireland2008}
{Ireland} M.~J.  et~al., 2008, in Society of Photo-Optical Instrumentation
  Engineers (SPIE) Conference Series.

\bibitem[\protect\citeauthoryear{{Kervella}, {Th{\'e}venin}, {Di Folco} \&
  {S{\'e}gransan}}{{Kervella} et~al.}{2004}]{Kervella2004}
{Kervella} P.,  {Th{\'e}venin} F.,  {Di Folco} E.,    {S{\'e}gransan} D.,
  2004, \aap, 426, 297

\bibitem[\protect\citeauthoryear{{Koleva} \& {Vazdekis}}{{Koleva} \&
  {Vazdekis}}{2012}]{Koleva2012}
{Koleva} M.,  {Vazdekis} A.,  2012, \aap, 538, A143

\bibitem[\protect\citeauthoryear{{Kornilov}, {Mironov} \&
  {Zakharov}}{{Kornilov} et~al.}{1996}]{Kornilov1996}
{Kornilov} V.,  {Mironov} A.,    {Zakharov} A.,  1996, Baltic Astronomy, 5, 379

\bibitem[\protect\citeauthoryear{{Lafrasse}, {Mella}, {Bonneau}, {Duvert},
  {Delfosse}, {Chesneau} \& {Chelli}}{{Lafrasse} et~al.}{2010}]{Lafrasse2010}
{Lafrasse} S.,  {Mella} G.,  {Bonneau} D.,  {Duvert} G.,  {Delfosse} X.,
  {Chesneau} O.,    {Chelli} A.,  2010, in Society of Photo-Optical
  Instrumentation Engineers (SPIE) Conference Series.

\bibitem[\protect\citeauthoryear{{Lane}, {Boden} \& {Kulkarni}}{{Lane}
  et~al.}{2001}]{Lane2001}
{Lane} B.~F.,  {Boden} A.~F.,    {Kulkarni} S.~R.,  2001, \apjl, 551, L81

\bibitem[\protect\citeauthoryear{{Maeder} \& {Meynet}}{{Maeder} \&
  {Meynet}}{2000}]{Maeder2000}
{Maeder} A.,  {Meynet} G.,  2000, \araa, 38, 143

\bibitem[\protect\citeauthoryear{{Maestro} et~al.,}{{Maestro}
  et~al.}{2012}]{Maestro2012}
{Maestro} V.  et~al., 2012, in Society of Photo-Optical Instrumentation
  Engineers (SPIE) Conference Series.

\bibitem[\protect\citeauthoryear{{Mason}, {Wycoff}, {Hartkopf}, {Douglass} \&
  {Worley}}{{Mason} et~al.}{2001}]{Mason2001}
{Mason} B.~D.,  {Wycoff} G.~L.,  {Hartkopf} W.~I.,  {Douglass} G.~G.,
  {Worley} C.~E.,  2001, \aj, 122, 3466

\bibitem[\protect\citeauthoryear{{Mermilliod}, {Mermilliod} \&
  {Hauck}}{{Mermilliod} et~al.}{1997}]{Mermilliod1997}
{Mermilliod} J.-C.,  {Mermilliod} M.,    {Hauck} B.,  1997, \aaps, 124, 349

\bibitem[\protect\citeauthoryear{{Monnier} et~al.,}{{Monnier}
  et~al.}{2012}]{Monnier2012}
{Monnier} J.~D.  et~al., 2012, \apjl, 761, L3

\bibitem[\protect\citeauthoryear{{Monnier} et~al.,}{{Monnier}
  et~al.}{2007}]{Monnier2007a}
{Monnier} J.~D.  et~al., 2007, Science, 317, 342

\bibitem[\protect\citeauthoryear{{Mozurkewich} et~al.,}{{Mozurkewich}
  et~al.}{2003}]{Mozurkewich2003}
{Mozurkewich} D.  et~al., 2003, \aj, 126, 2502

\bibitem[\protect\citeauthoryear{{Nieva} \& {Przybilla}}{{Nieva} \&
  {Przybilla}}{2012}]{Nieva2012}
{Nieva} M.-F.,  {Przybilla} N.,  2012, \aap, 539, A143

\bibitem[\protect\citeauthoryear{{Olevi{\'c}} \& {Cvetkovi{\'c}}}{{Olevi{\'c}}
  \& {Cvetkovi{\'c}}}{2006}]{Olevic2006}
{Olevi{\'c}} D.,  {Cvetkovi{\'c}} Z.,  2006, \aj, 131, 1721

\bibitem[\protect\citeauthoryear{{Peterson} et~al.,}{{Peterson}
  et~al.}{2006}]{Peterson2006}
{Peterson} D.~M.  et~al., 2006, \nat, 440, 896

\bibitem[\protect\citeauthoryear{{Pier}, {Saha} \& {Kinman}}{{Pier}
  et~al.}{2003}]{Pier2003}
{Pier} J.~R.,  {Saha} A.,    {Kinman} T.~D.,  2003, Information Bulletin on
  Variable Stars, 5459, 1

\bibitem[\protect\citeauthoryear{{Pourbaix} et~al.,}{{Pourbaix}
  et~al.}{2004}]{Pourbaix2004}
{Pourbaix} D.  et~al., 2004, \aap, 424, 727

\bibitem[\protect\citeauthoryear{{Prugniel}, {Vauglin} \& {Koleva}}{{Prugniel}
  et~al.}{2011}]{Prugniel2011}
{Prugniel} P.,  {Vauglin} I.,    {Koleva} M.,  2011, \aap, 531, A165

\bibitem[\protect\citeauthoryear{{Richichi}, {Percheron} \&
  {Khristoforova}}{{Richichi} et~al.}{2005}]{Richichi2005}
{Richichi} A.,  {Percheron} I.,    {Khristoforova} M.,  2005, \aap, 431, 773

\bibitem[\protect\citeauthoryear{{Royer}, {Zorec} \& {G{\'o}mez}}{{Royer}
  et~al.}{2007}]{Royer2007}
{Royer} F.,  {Zorec} J.,    {G{\'o}mez} A.~E.,  2007, \aap, 463, 671

\bibitem[\protect\citeauthoryear{{Smith} \& {Dworetsky}}{{Smith} \&
  {Dworetsky}}{1993}]{Smith1993}
{Smith} K.~C.,  {Dworetsky} M.~M.,  1993, \aap, 274, 335

\bibitem[\protect\citeauthoryear{{Tassoul}}{{Tassoul}}{2000}]{Tassoul2000}
{Tassoul} J.-L.,  2000, {Stellar Rotation}.
Cambridge University Press

\bibitem[\protect\citeauthoryear{{ten Brummelaar} et~al.,}{{ten Brummelaar}
  et~al.}{2005}]{tenBrummelaar2005}
{ten Brummelaar} T.~A.  et~al., 2005, \apj, 628, 453

\bibitem[\protect\citeauthoryear{{Tokovinin}}{{Tokovinin}}{1997}]{Tokovinin1997}
{Tokovinin} A.~A.,  1997, \aaps, 124, 75

\bibitem[\protect\citeauthoryear{{Underhill}, {Divan}, {Prevot-Burnichon} \&
  {Doazan}}{{Underhill} et~al.}{1979}]{Underhill1979}
{Underhill} A.~B.,  {Divan} L.,  {Prevot-Burnichon} M.-L.,    {Doazan} V.,
  1979, \mnras, 189, 601

\bibitem[\protect\citeauthoryear{{Vakili}, {Mourard}, {Bonneau}, {Morand} \&
  {Stee}}{{Vakili} et~al.}{1997}]{Vakili1997}
{Vakili} F.,  {Mourard} D.,  {Bonneau} D.,  {Morand} F.,    {Stee} P.,  1997,
  \aap, 323, 183

\bibitem[\protect\citeauthoryear{{van Belle}}{{van Belle}}{2012}]{vanBelle2012}
{van Belle} G.~T.,  2012, \aapr, 20, 51

\bibitem[\protect\citeauthoryear{{van Belle}, {Ciardi} \& {Boden}}{{van Belle}
  et~al.}{2007}]{vanBelle2007}
{van Belle} G.~T.,  {Ciardi} D.~R.,    {Boden} A.~F.,  2007, \apj, 657, 1058

\bibitem[\protect\citeauthoryear{{van Belle} et~al.,}{{van Belle}
  et~al.}{1999}]{vanBelle1999a}
{van Belle} G.~T.  et~al., 1999, \aj, 117, 521

\bibitem[\protect\citeauthoryear{{van Belle} \& {van Belle}}{{van Belle} \&
  {van Belle}}{2005}]{vanBelle2005}
{van Belle} G.~T.,  {van Belle} G.,  2005, \pasp, 117, 1263

\bibitem[\protect\citeauthoryear{{van Belle} \& {von Braun}}{{van Belle} \&
  {von Braun}}{2009}]{vanBelle2009}
{van Belle} G.~T.,  {von Braun} K.,  2009, \apj, 694, 1085

\bibitem[\protect\citeauthoryear{{van Leeuwen}}{{van
  Leeuwen}}{2007}]{vanLeeuwen2007}
{van Leeuwen} F.,  2007, \aap, 474, 653

\bibitem[\protect\citeauthoryear{{von Zeipel}}{{von
  Zeipel}}{1924a}]{vonZeipel1924a}
{von Zeipel} H.,  1924a, \mnras, 84, 665

\bibitem[\protect\citeauthoryear{{von Zeipel}}{{von
  Zeipel}}{1924b}]{vonZeipel1924b}
{von Zeipel} H.,  1924b, \mnras, 84, 684

\bibitem[\protect\citeauthoryear{{Wegner}}{{Wegner}}{2002}]{Wegner2002}
{Wegner} W.,  2002, Baltic Astronomy, 11, 1

\bibitem[\protect\citeauthoryear{{White} et~al.,}{{White}
  et~al.}{2013}]{White2013}
{White} T.~R.  et~al., 2013, \mnras

\bibitem[\protect\citeauthoryear{{Wu}, {Singh}, {Prugniel}, {Gupta} \&
  {Koleva}}{{Wu} et~al.}{2011}]{Wu2011}
{Wu} Y.,  {Singh} H.~P.,  {Prugniel} P.,  {Gupta} R.,    {Koleva} M.,  2011,
  \aap, 525, A71

\bibitem[\protect\citeauthoryear{{Yoon}, {Peterson}, {Armstrong}, {Clark} III,
  {Gilbreath}, {Pauls}, {Schmitt} \& {Zagarello}}{{Yoon}
  et~al.}{2007}]{Yoon2007}
{Yoon} J.,  {Peterson} D.~M.,  {Armstrong} J.~T.,  {Clark} III J.~H.,
  {Gilbreath} G.~C.,  {Pauls} T.,  {Schmitt} H.~R.,    {Zagarello} R.~J.,
  2007, \pasp, 119, 437

\bibitem[\protect\citeauthoryear{{Zhao} et~al.,}{{Zhao}
  et~al.}{2009}]{Zhao2009}
{Zhao} M.  et~al., 2009, \apj, 701, 209

\bibitem[\protect\citeauthoryear{{Zorec}, {Cidale}, {Arias}, {Fr{\'e}mat},
  {Muratore}, {Torres} \& {Martayan}}{{Zorec} et~al.}{2009}]{Zorec2009}
{Zorec} J.,  {Cidale} L.,  {Arias} M.~L.,  {Fr{\'e}mat} Y.,  {Muratore} M.~F.,
  {Torres} A.~F.,    {Martayan} C.,  2009, \aap, 501, 297

\end{thebibliography}
\bibliographystyle{mn2e}

\end{document}